\documentclass[12pt]{spieman}  % 12pt font required by SPIE;
\pdfoutput=1
\usepackage{amsmath,amsfonts,amssymb}
\usepackage{graphicx}
\usepackage{setspace}
\usepackage{tocloft}
\usepackage{xspace}
\usepackage{orcidlink}

\title{Mitigating the effects of particle background on the Athena Wide-Field Imager}
\author[a,*]{Eric D.\ Miller\,\orcidlink{0000-0002-3031-2326}}
\author[a]{Catherine E.\ Grant\,\orcidlink{0000-0002-4737-1373}} 
\author[a]{Marshall W.\ Bautz\,\orcidlink{0000-0002-1379-4482}}
\author[b]{Silvano Molendi\,\orcidlink{0000-0002-2483-278X}}
\author[c]{Ralph Kraft\,\orcidlink{0000-0002-0765-0511}}
\author[c,d]{Paul Nulsen\,\orcidlink{0000-0003-0297-4493}}
\author[c,e]{Esra Bulbul\,\orcidlink{0000-0002-7619-5399}}
\author[f]{Steven Allen}
\author[g]{David N.\ Burrows\,\orcidlink{0000-0003-0729-1632}}
\author[e]{Tanja Eraerds}
\author[h]{Valentina Fioretti\,\orcidlink{0000-0002-6082-5384}}
\author[b]{Fabio Gastaldello\,\orcidlink{/0000-0002-9112-0184}}
\author[i]{David Hall}
\author[i]{Michael W.J.\ Hubbard\,\orcidlink{0000-0002-4378-826X}}
\author[i]{Jonathan Keelan}
\author[e]{Norbert Meidinger\,\orcidlink{0000-0002-6756-413X}}
\author[j]{Emanuele Perinati}
\author[e]{Arne Rau\,\orcidlink{0000-0002-6756-413X}}
\author[f]{Dan Wilkins\,\orcidlink{0000-0002-4794-5998}}

\affil[a]{Kavli Institute for Astrophysics and Space Research, Massachusetts Institute of Technology, 77 Massachusetts Avenue, Cambridge, MA 02139, USA}
\affil[b]{INAF/IASF-Milano, Via Bassini 15, I-20133 Milano, Italy}
\affil[c]{Center for Astrophysics \textbar\ Harvard \& Smithsonian, 60 Garden Street, Cambridge, MA 02138, USA}
\affil[d]{ICRAR, University of Western Australia, 35 Stirling Highway, Crawley, WA 6009, Australia}
\affil[e]{Max Planck Institute for Extraterrestrial Physics, Giessenbachstr. 1 D-85748, Garching, Germany}
\affil[f]{Department of Physics, Stanford University, 382 Via Pueblo Mall, Stanford, CA 94305-4060, USA}
\affil[g]{Pennsylvania State University, Department of Astronomy and Astrophysics, 525 Davey Lab, University Park, PA 16802, USA}
\affil[h]{INAF Osservatorio di Astrofisica e Scienza dello Spazio di Bologna, via Gobetti 93/3, I-40129 Bologna, Italy}
\affil[i]{Centre for Electronic Imaging, The Open University, Walton Hall, Milton Keynes, MK7 6AA, UK}
\affil[j]{Institut f\"{u}r Astronomie und Astrophysik, Universit\"{a}t T\"{u}bingen, Sand 1, 72076 T\"{u}bingen, Germany}

\newcommand\procspie{Proc.~SPIE}%      % Proceedings of the SPIE 
\newcommand\aj{Astron.~J.}%    % Astronomical Journal ++
\newcommand\apj{Astrophys.~J.}%    % Astrophysical Journal ++
\newcommand\aap{A\&A}%    % Astronomy & Astrophysics
\newcommand\jatis{JATIS}%    % Astronomy & Astrophysics
\newcommand\pcor{$P_{cor}$}

\cftpagenumbersoff{figure}
\cftpagenumbersoff{table} 

\begin{document} 
\maketitle

\begin{abstract}
The Wide Field Imager (WFI) flying on Athena will usher in the next era of studying the hot and energetic Universe. Among Athena's ambitious science programs are observations of faint, diffuse sources limited by statistical and systematic uncertainty in the background produced by high-energy cosmic ray particles. These particles produce easily identified ``cosmic-ray tracks'' along with less easily identified signals produced by secondary photons or X-rays generated by particle interactions with the instrument. Such secondaries produce identical signals to the X-rays focused by the optics, and cannot be filtered without also eliminating these precious photons. As part of a larger effort to estimate the level of unrejected background and mitigate its effects, we here present results from a study of background-reduction techniques that exploit the spatial correlation between cosmic-ray particle tracks and secondary events. We use Geant4 simulations to generate a realistic particle background signal, sort this into simulated WFI frames, and process those frames in a similar way to the expected flight and ground software to produce a realistic WFI observation containing only particle background. The technique under study, Self Anti-Coincidence or SAC, then selectively filters regions of the detector around particle tracks, turning the WFI into its own anti-coincidence detector. We show that SAC is effective at improving the systematic uncertainty for observations of faint, diffuse sources, but at the cost of statistical uncertainty due to a reduction in signal. If sufficient pixel pulse-height information is telemetered to the ground for each frame, then this technique can be applied selectively based on the science goals, providing flexibility without affecting the data quality for other science. The results presented here are relevant for any future silicon-based pixelated X-ray imaging detector, and could allow the WFI and similar instruments to probe to truly faint X-ray surface brightness.
\end{abstract}

% Include a list of up to six keywords after the abstract

% Include email contact information for corresponding author
{\noindent \footnotesize\textbf{*}Send all correspondence to Eric D.\ Miller, \linkable{milleric@mit.edu}}

%%%%%%%%%%%%%%%%%%%%%%%%%%%%%%%%%%%%%%%%%%%%%%%%%%%%%%%%%%%%%%%
% Introduction
%%%%%%%%%%%%%%%%%%%%%%%%%%%%%%%%%%%%%%%%%%%%%%%%%%%%%%%%%%%%%%%
\section{Introduction}
\label{sect:intro}

Silicon-based X-ray imaging instruments typically characterize detected photons by reconstructing the energy from the spatial pattern of electrons liberated by the photon interaction with the detector substrate. This technique, while allowing the detector to be used as an imaging spectrometer, is complicated by the fact that highly energetic charged particles undergo similar interactions in such detectors, producing signals that can be difficult to separate from the photon signal produced by a celestial source and properly focused by the optics. In attempts to detect extended, very low surface brightness sources such as galaxy cluster outskirts and the Warm Hot Intergalactic Medium, this cosmic-ray-induced background is the dominant source of both statistical and systematic error, the latter arising from our incomplete knowledge of the time and spectral variability of the underlying particle flux. 

Understanding and minimizing this particle background is vital for future advanced X-ray imagers, which will attempt to detect this faint extended emission in long exposures dominated by signals from cosmic-ray protons, alpha particles, and electrons, as well as photons from the Galactic foreground and extragalactic background. The Wide Field Imager (WFI)\cite{Meidinger2017} to fly on Athena\cite{AthenaWP2013}, ESA's next large X-ray observatory, is one such instrument. It will fly a 40$^\prime$ field-of-view array of DEPFET (depleted p-channel field-effect transistor) active pixel sensors, fully depleted to 450 $\mu$m with a pixel size of 130$\times$130 $\mu$m, and operating in the 0.2--15 keV band with a full-frame readout time of 5 ms. The Athena science requirements for the non-X-ray background are a count rate less than $5.5\times10^{-3}$ counts\,s$^{-1}$\,cm$^{-2}$\,keV$^{-1}$ in the 2--7 keV band and knowledge of the background to within a few percent\cite{AthenaSciReq2.6}, both challenging goals for a silicon detector in orbit at either L1 or L2. These requirements, based on the ambitious faint-source science goals, require careful pre-launch work to both predict the level of background and develop algorithms to reduce and characterize it once in orbit.

Previous generations of X-ray detectors have generally used one of two methods to reduce background from cosmic-ray particles: (1) identifying and eliminating events with pixel activation patterns more likely to be associated with particle tracks than with X-ray photons; or (2) the use of anti-coincidence detectors positioned close to the science detector enabling simultaneous detection of particle tracks and dropping of events when a signal appears in both detectors.  Strategy (1) is useful in eliminating events produced by the primary particle itself, but such particles can produce secondaries when interacting with the instrument structure. Secondaries that are low-energy photons or electrons have indistinguishable pixel patterns from the cosmic X-rays constituting the signal, and thus there is an irreducible limit to how well the background can be rejected by simply considering the event shape.  Strategy (2) overcomes this obstacle by eliminating all signal recorded during the primary particle interaction, including secondaries. However, for non-triggered detectors, if the integration time is comparable to the expected arrival interval of cosmic-ray primaries, then most of the frames will be rejected and much of the real signal will be lost. 

Due to its particular characteristics of detector size, pixel size, and especially its 5-ms frame time, the WFI inhabits a realm where both of these methods have some strength, and in the end the choice made depends sensitively on the science goals of an observation. Since the pattern-based background rejection technique has been employed on several previous and operating missions, including XMM-Newton EPIC, Chandra ACIS, Swift XRT, and Suzaku XIS, it is useful to analyze this real-world data. These instruments have the benefits that we understand their design and function well, and for some we have a large amount of full-frame data which contains information from all pixels, including particle tracks. However, the detectors are different in design and operation from the WFI DEPFETs, especially ACIS and XIS, and Swift and Suzaku are additionally in low-Earth orbit, a very different particle environment from Chandra and XMM-Newton in high-Earth orbit and the expected L1 or L2 orbit of Athena. This analysis is nevertheless illuminating, as we found strong spatial and temporal correlations between particle tracks produced by high-energy cosmic rays and events that would be interpreted as source X-rays. 
\cite{Grantetal2018,Bulbuletal2018,Bulbuletal2020}.

A large effort has been underway for several years to predict and model the expected WFI particle background using Geant4\cite{Geant4,Geant4_paper2} simulations, and to use these simulations to inform the design of both the camera shielding and on-board event filtering.\cite{vonKienlin2018,Grantetal2020,Eraerdsetal2020,Eraerdsetal2021}  In this work, we use a set of these Geant4 simulations of cosmic rays interacting with the WFI camera body to model the expected unrejected particle background and explore techniques to separate this signal from the desired X-ray signal. In particular, we study correlations between those unrejected events and cosmic ray tracks produced by the same primary particle interaction; these latter signals have historically been eliminated from telemetered data due to bandwidth constraints. As we show, there is a direct spatial correlation between particle tracks and apparently valid events that can be exploited to, in effect, use the WFI as its own anti-coincidence detector and reduce the unrejected particle background in a statistical sense. This ``Self Anti-coincidence'' (SAC) method exploits both the spatial correlation between particle tracks and valid events, and the particular frame time of the WFI, during which we expect an average of a few cosmic ray interactions that produce signal in the detector. We present results from this analysis along with a description of how SAC can be tuned depending on the science goals of a particular observation. This technique is applicable to any future astronomical X-ray imaging instrument with a fast frame rate, provided sufficient information is telemetered for each frame.

This paper is organized as follows. In Section \ref{sect:data}, we describe the Geant4 simulation output and how this was converted into simulated WFI frames and event lists, along with characteristics of the simulated background signal and validation based on existing XMM-Newton data. In Section \ref{sect:results}, we present the results of an analysis of the spatial correlation of particle tracks and unrejected, X-ray-like events, along with an application and exploration of the SAC technique. In Section \ref{sect:summary} we summarize our findings. An explanation of SAC and its various metrics of background reduction as developed by the WFI Background Working Group (BWG) are presented in Appendix \ref{sect:app_sac}.

%%%%%%%%%%%%%%%%%%%%%%%%%%%%%%%%%%%%%%%%%%%%%%%%%%%%%%%%%%%%%%%
% Data and Analysis
%%%%%%%%%%%%%%%%%%%%%%%%%%%%%%%%%%%%%%%%%%%%%%%%%%%%%%%%%%%%%%%
\section{Data \& analysis}
\label{sect:data}

\subsection{Geant4 simulations and sorting of data}
\label{sect:geant}

The Geant4 simulations were performed at The Open University and consisted of 133 runs of $10^6$ Galactic cosmic ray (GCR) proton primaries per run, drawn from the CREME 96 standard spectral model for solar minimum \cite{Tylka1997} and generated on a 70-cm radius sphere surrounding the WFI instrument. These simulations used a simplified WFI mass model designated E0015261, which includes the camera, proton shield, filter wheel, and baffle, but excludes a graded-Z shield under later study by the WFI BWG to reduce the impact of energetic cosmic X-ray background photons and of secondary electrons produced by GCR interactions in the proton shield. This is the same mass model used to obtain results previously presented,\cite{vonKienlin2018} and we refer the reader there for more detailed information about the Geant4 simulation setup and operation. For each GCR primary that generated signal charge in the WFI detector, the data include the deposited energy in keV in each pixel and information about the particle (primary or secondary) responsible for the deposition.  The vast majority of simulated primaries do not interact with the WFI detector; indeed, only 936,934 of 133,000,000 (0.7\%) produce signal in any pixels. 

The Geant4 output was structured into two different formats for further analysis.  The first dataset was structured on a primary-by-primary basis, hereafter referred to as ``single-primary'' frames, and this was used to explore fundamental properties of the signal produced by individual cosmic rays and search for useful correlations between particle tracks and events that look like X-rays that could be exploited to flag the latter. The second type of dataset has primary GCRs randomly sorted into frames of a finite exposure time to simulate a real-world observation of the WFI background.  While the WFI is expected to operate at 5 ms per frame\cite{Meidinger2017},  we simulated a range of frame times from 0.2 ms to 5 ms, and focus here specifically on 5 ms and 2 ms, to compare the effects of readout rate on SAC background reduction.  Considering different frame times also serves as a proxy for sampling solar cycle variability, since a 2-ms frame will have 40\% of the particle fluence of a 5-ms frame, similar to the factor of $\sim$2 difference in GCR flux observed between solar maximum and minimum\cite{Grantetal2018}. To construct the datasets, we sorted primaries into frames using the effective total exposure time given by Eq.\,4 of Fioretti et al.~(2012)\cite{Fiorettietal2012},
\begin{equation}
t_{exp} = \frac{N_p}{\Phi \times 4 \pi^2 R^2} = \frac{N_p}{\phi\,\pi R^2},
\end{equation}	
where $N_p$ is the number of simulated primary protons, $\Phi$ is the cosmic ray proton intensity in units of cm$^{-2}$\,s$^{-1}$\,sr$^{-1}$ at the assumed Athena L1 or L2 orbit, $\phi = 4\pi\Phi$ is the cosmic ray proton flux in units of cm$^{-2}$\,s$^{-1}$,
and $R = 70$ cm is the radius of the simulation boundary sphere. The conversion from intensity to flux assumes an isotropic cosmic ray intensity, and like Fioretti et al.~(2012)\cite{Fiorettietal2012}, we have drawn simulated protons from a cosine law angular distribution, although without restricting the flux to a small cone. 

We assume $\phi$ = 4.1 cm$^{-2}$\,s$^{-1}$ for GCR protons, based on SPENVIS\cite{SPENVIS} simulations of the CREME 96 spectral model for solar minimum \cite{Tylka1997}, yielding $t_{exp} = 15.8$ s for a single Geant4 run of $N_p = 10^6$ primaries. As we show below, this proton flux produces an average 2--7 keV unrejected count rate consistent with that derived previously by the WFI BWG for protons only, $5\times10^{-3}$ cm$^{-2}$\,s$^{-1}$\,keV$^{-1}$ \cite{vonKienlin2018}. However, since the real particle background environment includes other species such as GCR alpha particles, electrons, and gamma rays, we increased the proton flux by 40\% to account for these primaries missing from the Geant4 simulations. This produced a total average 2--7 keV unrejected count rate consistent with that found by previous Geant4 analysis amongst the BWG\cite{vonKienlin2018}, $\sim7\times10^{-3}$ cm$^{-2}$\,s$^{-1}$\,keV$^{-1}$. We note that the details of the secondary interactions are likely different between protons and these other species, but to first order this is a reasonable approximation. We also note that this is a reasonable upper limit to the GCR flux, as it is based on recent solar minimum observations and in an extended mission Athena could observe during all parts of one or more solar cycles.

The scaled GCR primary flux yields a total effective exposure time of 1505 s for the 133 million primaries, a rate of $8.84\times10^{4}$ s$^{-1}$, or 441.9 per 5-ms frame (176.8 per 2-ms frame). Using this as the mean rate, each of the 133 million primaries was assigned a random arrival time drawn from an exponential distribution, appropriate for modeling arrival intervals of this Poisson process. Primaries were then assigned into each frame according to these arrival times.  We determine a mean rate of 3.11 interacting primaries per frame in the 300,967 5-ms frames that were simulated. Of these frames, 95.5\% have signal in them, consistent with the expectation from the assumed Poisson distribution. The simulated 2-ms frames are similarly consistent, with an average rate of 1.25 interacting primaries per frame, and 71.2\% of the 752,331 total frames containing signal.

For each case (single-primary, 5-ms, and 2-ms frames), each frame with signal was turned into an image of pixel values using the pixel X, Y, and deposited energy information provided by Geant4. These simulations recorded signal deposited in a $1181\times1181$ pixel grid, using 130-$\mu$m pixels and including a 3-pixel (0.42-$\mu$m) gap between the quadrants. This is larger than the full WFI Large Detector Array (LDA) field of view, with $512\times512$ pixel quadrants, or a $1027\times1027$ pixel full field including the same gaps. While assembling frames, we simply excised the outer pixels. Any primaries that have signal only in the excised region were treated as though they had not interacted with the detector. Any primaries that had signal in both the outer (excised) and inner regions had their outer signal removed and inner signal retained. We note that this chip gap is significantly smaller than the likely WFI design gap, $\sim4$ mm.

\subsection{Identifying valid events and particle tracks}
\label{sect:event_finding}

\begin{figure}[t]
\begin{center}
\includegraphics[width=6.25in]{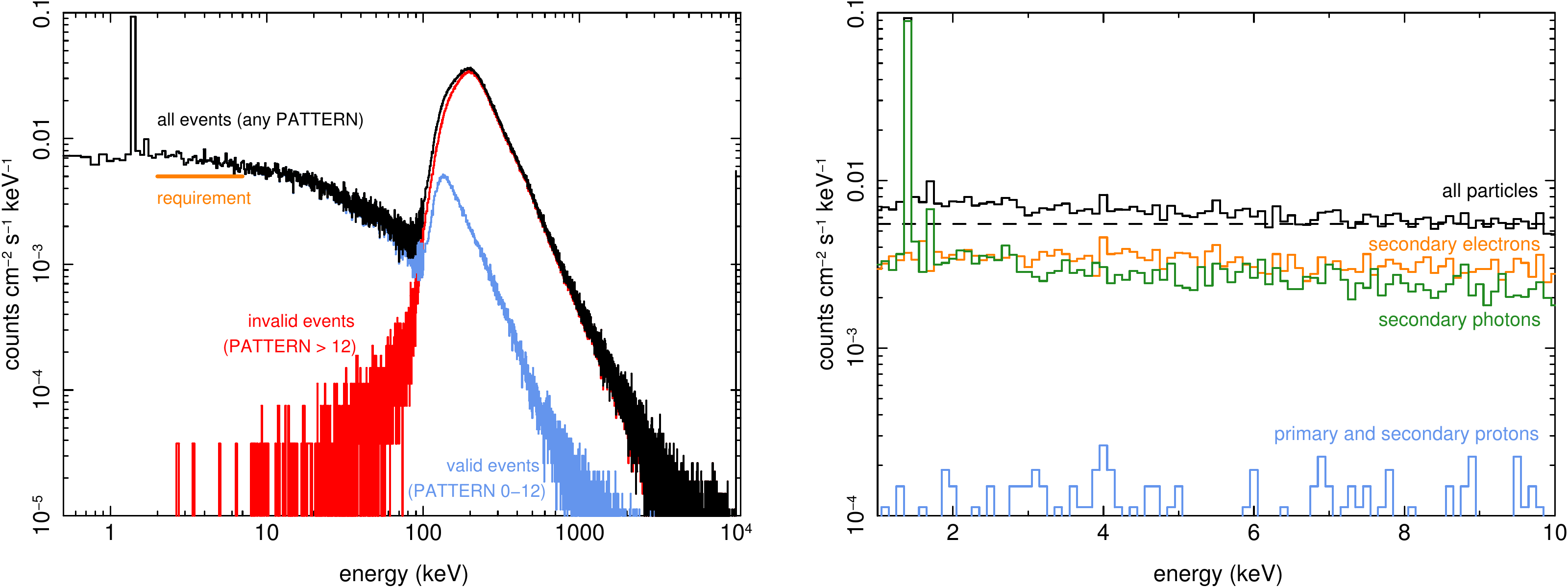}
\caption{Spectra of events produced by the Geant4 GCR proton primary simulations. (left) The spectrum over a wide energy band, showing pattern-based valid and invalid events separately. Valid events dominate by several orders of magnitude in the 2--7 keV band, while invalid events dominate above 100 keV, well outside the WFI sensitive band. (right) Spectrum in the 2--7 keV region, with the WFI unrejected background requirement of $5.5\times10^{-3}$ s$^{-1}$ cm$^{-2}$ keV$^{-1}$ plotted as a dashed line. Colored lines indicate what types of particles produce the detected signal for these events--primarily secondary electrons and photons produced in primary proton interactions with the WFI. The strong line near 1.5 keV is Al K$\alpha$, and the weaker line near 1.7 keV is Si K$\alpha$.}
\label{fig:event_spectra}
\end{center}
\end{figure}

Each image was searched for events using a local-maximum method similar to that employed on-board many X-ray imaging instruments like XMM-Newton EPIC pn and Chandra ACIS. First an event threshold of 0.1 keV was applied, and pixels at or above this level were flagged as event candidates. Each candidate pixel was compared to the other pixels in its 3$\times$3 pixel neighborhood, and if it was a local maximum it was flagged as an event center. The 5$\times$5 neighborhood around each event center was then searched for pixels at or above the neighbor (or split) threshold, also set at 0.1 keV. The event pattern was assigned using EPIC pn rules\cite{XMMSAS}, including single-pixel events (PATTERN=0), doubles (PATTERN=1--4), triples (PATTERN=5--8), quadruples (PATTERN=9--12), and everything else (PATTERN=13). In particular, for all non-single-pixel events which have a 3$\times$3 neighbor above the neighbor threshold, the outer 5$\times$5 was also searched for pixels above the neighbor threshold. Double, triple, and quad patterns with at least one outer 5x5 pixel above the neighbor threshold were assigned PATTERN=13. In the remainder of this work, ``valid'' events (used interchangeably with ``counts'') are those with PATTERN$<$13, as these are indistinguishable from events produced by X-ray photons. The energy of the event is the summed energy of all pixels in the inner 3$\times$3 island that are above the neighbor threshold. Because of the 5$\times$5 pattern assignment, events with centers within 2 pixels of the edge of a quadrant were excluded. This reduces the sensitive detector area by 1.6\%. Figure \ref{fig:event_spectra} shows the spectra of valid, invalid, and all events.

\begin{figure}[p]
\begin{center}
\includegraphics[width=3.0in]{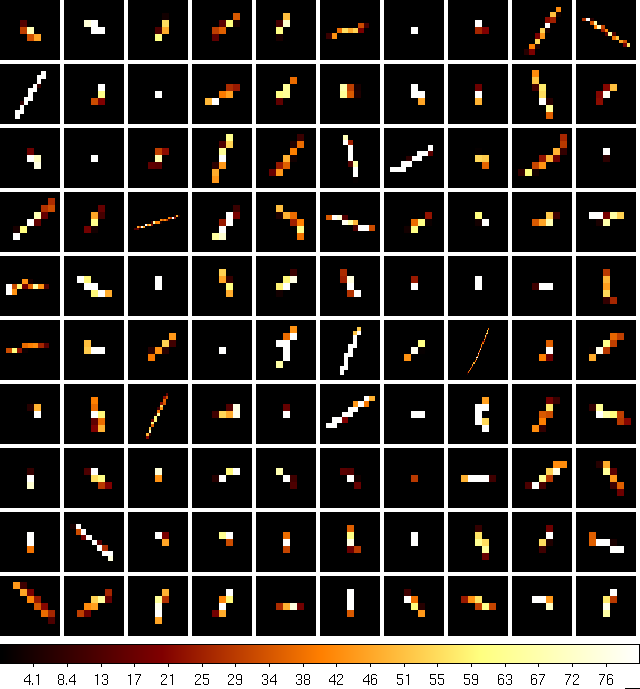}
\caption{Images of a small sample of individual particle tracks, with the color scale in keV. Pixels shown are equivalent to 130 $\mu$m WFI LDA pixels, so the image sizes are not the same and scale with the size of the tracks.}
\label{fig:blob_pstamps}
\end{center}
\end{figure}

We identified particle tracks using image segmentation in each frame. Hereafter, a  ``particle track'' is defined as a pattern which is either (1) a spatially contiguous set of five or more pixels above the neighbor threshold, 0.1 keV; or (2) any contiguous set of pixels above 0.1 keV that includes at least one pixel over 15 keV. This latter energy is called the ``MIP threshold'', an energy above which the Athena mirrors have effectively zero efficiency, and thus all signal is assumed to be produced by cosmic ray minimum ionizing particles, or ``MIPs''. Detached diagonals are considered contiguous in this image segmentation, and we did not apply the spatial edge filtering to particle tracks as we did to events, since these regions contain useful knowledge about their presence. Note that our definition of ``particle track'' differs slightly from that used for the EPIC pn analysis\cite{Bulbuletal2020} due to option (2). Each particle track was assigned an ID number to uniquely identify it in the full dataset. Examples of particle tracks are shown as postage stamps in Figure \ref{fig:blob_pstamps}. A single primary can produce multiple detached particle tracks.

\begin{figure}[p]
\begin{center}
\includegraphics[width=3.0in]{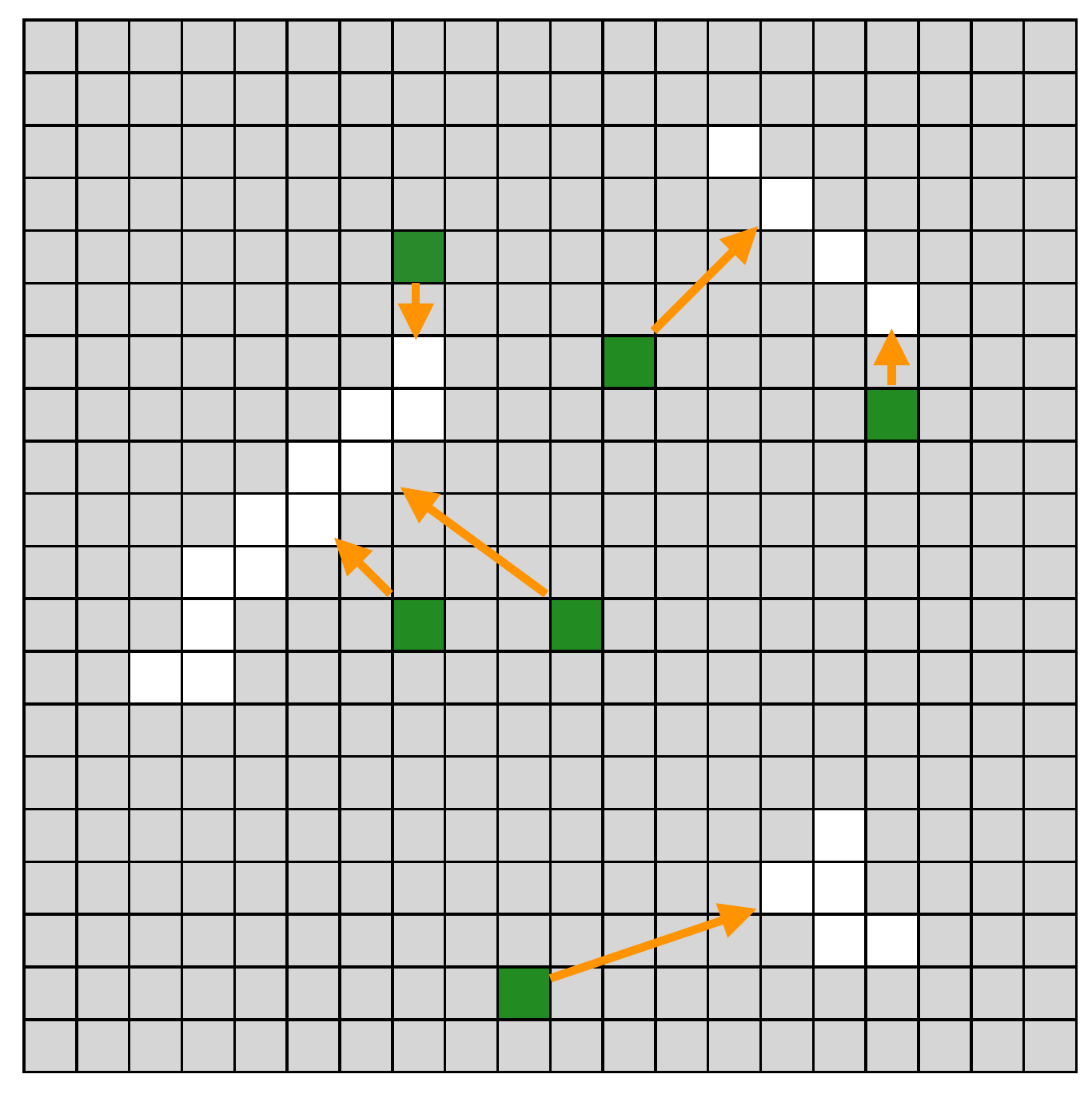}
\caption{Schematic of a frame containing particle tracks (white pixels) and valid events (green pixels). The image segmentation would identify three particle tracks in this frame. Orange arrows indicate the distance between each valid event, defined by the maximum pixel of the 3$\times$3 island, and the nearest particle track or MIP pixel.}
\label{fig:distance_scheme}
\end{center}
\end{figure}

Finally, in each frame, the distance between the central pixel of each event and the nearest pixel in a particle track was calculated. Many events fall on particle tracks and so have a distance of zero. Valid events are by definition unable to fall on a particle track pixel. Thus valid events and particle tracks are a mutually exclusive set of entities, despite the different methods used to identify them. A schematic diagram of this distance finding technique is shown in Figure \ref{fig:distance_scheme}.

To aid our analysis of the correlations between particle tracks and valid events, we assigned frames to ``cases'' in the same way as the XMM-Newton EPIC pn analysis\cite{Bulbuletal2020}, namely:
\begin{itemize}
    \item Case A: frame contains only particle tracks.
    \item Case B: frame contains only valid events.
    \item Case C: frame contains both particle tracks and valid events.
    \item Case D: frame contains neither particle tracks nor valid events (empty frame).
\end{itemize}
This sorting was done for the single-primary frames as well as the 2-ms and 5-ms frames. Summary information about the fraction of frames and rates of particle tracks and valid events in each case is given in Table \ref{tab:frame_summary} and explored in more detailed in the following sections.  

\begin{table}[t]
\begin{center}
\caption{Summary information for frame-by-frame analysis.}\label{tab:frame_summary}
\begin{tabular}{lrrr}\hline \hline
Type of Frame                    &  Single primary  &  5 msec           &  2 msec           \\
\hline                                                                   
~no.~frames                      &  133,000,000     &  300,967          &  752,331          \\
~no.~frames with signal          &  936,934 (0.7\%) &  287,424 (95.5\%) &  535,360 (71.2\%) \\
~no.~frames with particle track  &  918,662 (0.7\%) &  286,580 (95.2\%) &  530,114 (70.5\%) \\
~no.~particle tracks per frame   &  0.0078          &  3.45             &  1.38             \\
\hline
\multicolumn{4}{l}{Case A (frame with only particle tracks)} \\
~fraction of all frames          &  0.68\%          &  87.3\%           &  67.8\%           \\
~fraction of frames with signal  &  97.3\%          &  91.4\%           &  95.3\%           \\
~no.~particle tracks per frame   &  1.12            &  3.57             &  1.94             \\
~fraction of valid events        &  ~$\cdots$~      &  ~$\cdots$~       &  ~$\cdots$~       \\
\hline
\multicolumn{4}{l}{Case B (frame with only valid events)} \\
~fraction of all frames          &  0.013\%         &  0.3\%            &  0.6\%            \\
~fraction of frames with signal  &  1.8\%           &  0.3\%            &  0.9\%            \\
~no.~particle tracks per frame   &  ~$\cdots$~      &  ~$\cdots$~       &  ~$\cdots$~       \\
~fraction of valid events        &  64.9\%          &  3.1\%            &  18.8\%           \\
\hline
\multicolumn{4}{l}{Case C (frame with both particle tracks and valid events)} \\
~fraction of all frames          &  0.007\%         &  7.9\%            &  2.7\%            \\
~fraction of frames with signal  &  0.9\%           &  8.3\%            &  3.8\%            \\
~no.~particle tracks per frame   &  1.89            &  4.20             &  2.51             \\
~fraction of valid events        &  35.1\%          &  96.9\%           &  81.2\%           \\
\hline
\multicolumn{4}{l}{Case D (frame with neither particle tracks nor valid events)} \\
~fraction of all frames          &  99.3\%          &  4.5\%            &  28.9\%           \\
~fraction of frames with signal  &  ~$\cdots$~      &  ~$\cdots$~       &  ~$\cdots$~       \\
~no.~particle tracks per frame   &  ~$\cdots$~      &  ~$\cdots$~       &  ~$\cdots$~       \\
~fraction of valid events        &  ~$\cdots$~      &  ~$\cdots$~       &  ~$\cdots$~       \\
\hline
\end{tabular}
\end{center}
\end{table}

\subsection{Validating the Geant4 Simulations}
\label{sect:validation}

The XMM-Newton EPIC pn data\cite{Bulbuletal2020} were used to validate the Geant4 simulations to give us confidence that the latter represent a reasonable simulation of the expected WFI particle environment and background. To first order, the relative fractions of Case A, B, and C frames should be similar between the two, although there are differences in the instruments. These include frame time (5.7 ms for EPIC pn vs.~5 ms for WFI), pixel size (150 $\mu$m for EPIC pn vs.~130 $\mu$m for WFI), and depletion depth (280 $\mu$m for EPIC pn vs.~450 $\mu$m for WFI), along with the absence of detector effects like charge splitting in the WFI simulations, differences in the instrument structure and shielding, and differences in the particle environment in the XMM-Newton high-Earth orbit and the Athena orbit at L1 or L2.

To mimic the Small Window Mode (SWM) used in the EPIC pn study, we used the 5 ms frames from Geant4 and, in every frame, drew a 64$\times$64 pixel square that included a pixel randomly chosen from those pixels with signal above the lower threshold. The center of the square was randomly assigned as long as it contained that pixel and fell within the limits of the full LDA field of view. Event finding and particle track image segmentation were performed using only the pixels within this square, using the methods described in Section \ref{sect:event_finding}. In this way we performed a similar processing to the EPIC pn SWM mode data, but only including frames with signal. These frames were sorted into Cases A, B, and C, and their relative fractions are shown in Table \ref{tab:validation}. Of the frames with signal, half as many (1\% vs. 2\%) contain valid events in the Geant4 data compared to the EPIC pn SWM data. Of these frames with valid events (Case B and C), we find that 86\% also have a particle track in the Geant4 analysis, virtually identical to the 87\% value for the EPIC pn SWM frames. That there is a higher fraction of frames with valid events in the XMM data could reflect differences in instrument design and operation, or an underestimation of the background rate from Geant4 similar to what has been seen on eROSITA \cite{Freybergetal2020}. The similarity of the Case B and C fractions is remarkable, however, and we conclude that the Geant4 simulations produce a valid representation of the expected WFI background for our purposes of exploring correlations between particles tracks and valid events.

\begin{table}[t]
\begin{center}
\caption{Relative frequency of different frame cases in EPIC pn data and Geant4 simulations.}\label{tab:validation}
\begin{tabular}{lcccc}\hline \hline
Frame Type                             &  \multicolumn{2}{l}{Fraction of Frames with}  &  \multicolumn{2}{l}{Fraction of Frames with}  \\
                                       &  \multicolumn{2}{l}{Signal (Case A+B+C)}      &  \multicolumn{2}{l}{Valid Events (Case B+C)}  \\
\hline
                                       &  EPIC-pn  &  Geant4                           &  EPIC-pn  &  Geant4  \\
\hline
Case A (particle track only)           &  98.00\%  &  99.04\%                          &  ~$\cdots$~  &  ~$\cdots$~ \\
Case B (valid event only)              &  1.75\%   &  0.83\%                           &  87\%        &  86\%      \\
Case C (particle track + valid event)  &  0.25\%   &  0.13\%                           &  13\%        &  14\%      \\
\hline
\end{tabular}
\end{center}
\end{table}

\subsection{Spectral Properties of Particle Tracks in Geant4 Simulations}
\label{sect:spectra}

\begin{figure}[t]
\begin{center}
\includegraphics[width=6.25in]{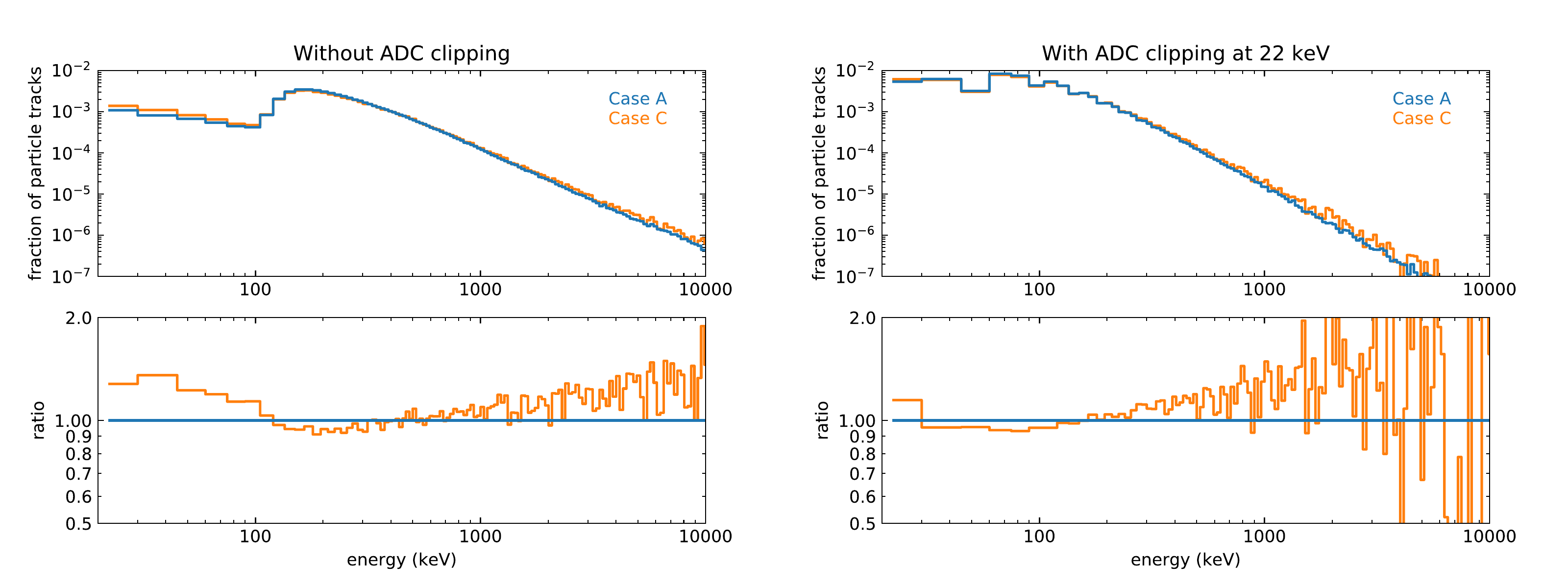}
\caption{Spectra of summed particle track energy. The left plot sums full pixel values, while the right plot clips pixels at 22.5 keV to mimic the dynamic range of EPIC pn, similar to that expected for the WFI. The top plots show both Case A and Case C, while the bottom plots normalize by Case A. A high-energy excess is seen in Case C particle tracks, similar to what is seen in EPIC pn\cite{Bulbuletal2020}. It is seen in both unclipped and clipped spectra.}
\label{fig:spectral_excess}
\end{center}
\end{figure}

With all particle tracks identified, we explored whether the small but systematic differences seen in the EPIC pn SWM Case A and Case C particle track spectra data\cite{Bulbuletal2020} were also present in the Geant4 simulation data. For each particle track, we calculated two versions of the total energy, first using the full range of pixel energies, and second clipping each pixel at 22.5 keV to mimic the dynamic range of EPIC pn, similar to that expected for the WFI. In both case, the pixels in a particle track are then summed to get the total energy, the distribution of which is shown in Figure \ref{fig:spectral_excess}. Interestingly, we see a flattening or excess of Case C particle tracks at high energy compared to Case A, similar to what is seen in the EPIC pn SWM data and providing further validation that the Geant4 results produce a reasonable simulation of the background. This may indicate different secondary particle production mechanisms for the Case C particle tracks, which are accompanied by valid events, compared to the Case A particle tracks, which are not. The fact that this difference is also seen in the clipped pixel data suggests a possible method of using the particle track energies to statistically identify frames which are likely to contain unrejected background. The level of background improvement and feasibility of this method are left for future work.

%%%%%%%%%%%%%%%%%%%%%%%%%%%%%%%%%%%%%%%%%%%%%%%%%%%%%%%%%%%%%%%
% Results
%%%%%%%%%%%%%%%%%%%%%%%%%%%%%%%%%%%%%%%%%%%%%%%%%%%%%%%%%%%%%%%
\section{Results}
\label{sect:results}

\subsection{Self-Anti-Coincidence (SAC)}
\label{sect:sac}

That valid events are spatially correlated with primary or secondary particle tracks from the same interacting cosmic ray was recognized early on in Geant4 simulations by the WFI Background Working Group\cite{vonKienlin2018} and in the analysis of in-orbit Chandra and Swift data\cite{Grantetal2018,Bulbuletal2018}. This correlation can be exploited by masking around particle tracks and flagging valid events within a certain distance; such events can later be filtered in ground processing depending on the science goals of the observation. However, this masking also reduces the signal and thus the efficiency of the instrument. This optional, partial-veto method has been termed ``Self-Anti-Coincidence'' (SAC), since under this scheme the WFI detector acts as its own anti-coincidence detector. Throughout the remainder of this work, we analyze the effects of SAC on different background reduction metrics, and explore the background improvement possible with enhanced, SAC-enabled post-processing algorithms.

\begin{figure}[t]
\begin{center}
\includegraphics[width=6.25in]{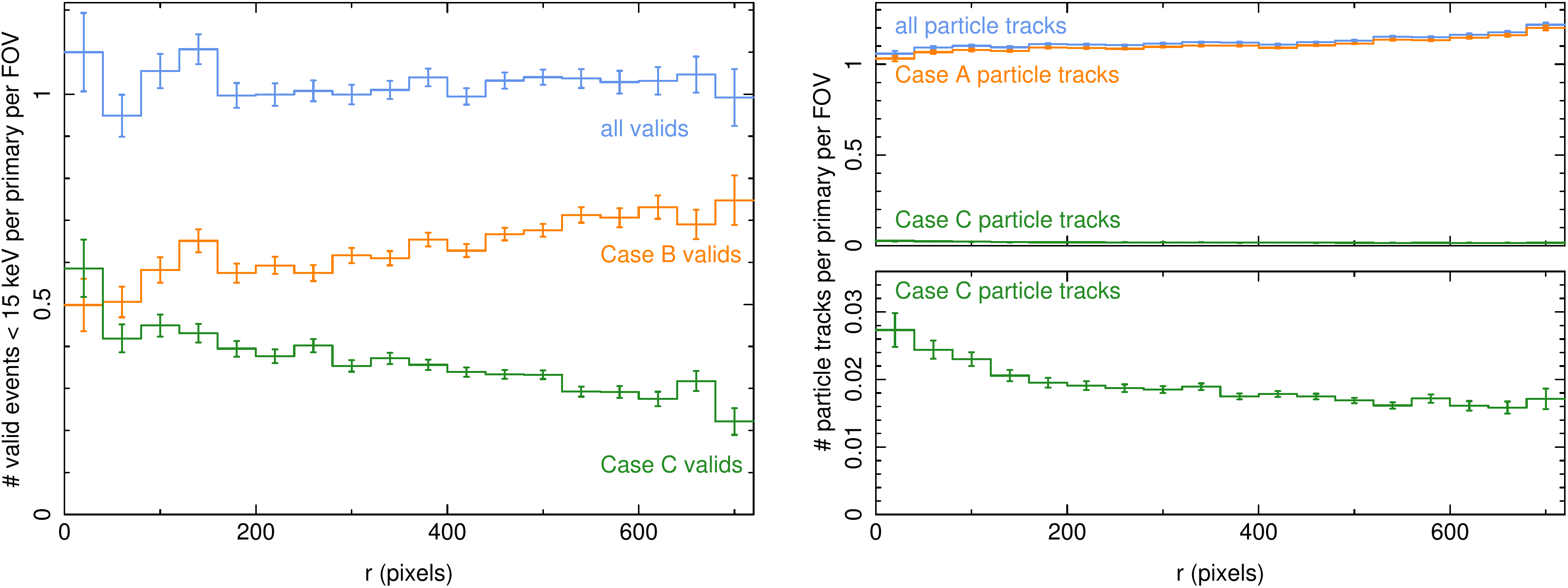}
\caption{Radial distribution of valid events (left) and particle tracks (right), normalized to the detector area. The lower panel of the right plot is a zoom-in to more clearly show the Case C points. The valid events overall have a flat distribution, however those valid events that accompany a particle track (Case C) are concentrated toward the center, and those that have no particle track (Case B) are more likely to be found near the edge. The particle tracks for those cases follow similar trends. This is expected, as a valid event detected near the edge is more likely to lose an accompanying particle track outside the field of view.}
\label{fig:raddist}
\end{center}
\end{figure}

\subsection{The Empirical Correlation Between Particle Tracks and Valid Events}
\label{sect:singleprim}

Frames containing single cosmic ray primary particles are key to understanding the spatial correlation between particle tracks and valid events. The area-normalized radial distributions of valid events and particle tracks derived from these single-primary frames are shown in Figure \ref{fig:raddist}. While the valid events have a flat distribution overall, those that accompany particle tracks (Case C) are more likely to be found toward the center of the frame, and those that lack a particle track (Case B) are more likely near the edge. The particle tracks for those cases follow similar trends. This is expected, since a valid event detected near the edge is more likely to lose an accompanying particle track off the edge.  

A useful metric to quantify this spatial correlation is the cumulative probability that a valid event falls within a certain radius of a particle track resulting from the same cosmic ray interaction.  We define this probability as \pcor$(<r_e)$, where $r_e$ is the ``exclusion radius'' to indicate its use in filtering unrejected background. A detailed analytic derivation of \pcor\, is presented in Appendix \ref{sect:app_pcor}, based on results from a previously published WFI Geant4 study\cite{vonKienlin2018}. We determine \pcor\, empirically from our Geant4 results as the cumulative distribution of radius in pixels between all Case C valid events and the nearest pixel in a particle track (the orange vectors in Figure \ref{fig:distance_scheme}).  To normalize \pcor\, to the full LDA field of view, we assume that Case B valid events have a corresponding particle track somewhere outside of the field. Thus we divide the distribution by the total number of valid events in Cases B and C. The resulting distribution is shown in Figure \ref{fig:pcor}, plotted with the analytic \pcor\, curves from Figure \ref{fig:prob_cor} in Appendix \ref{sect:app_pcor}, with lines for an infinite plane (black), a full LDA field (blue), and an LDA quadrant (red). Our orange curve is consistent with the model for the full LDA field, despite the very different methods used to derive the two. At the largest $r_e$, the correlation probability reaches 35\%. This is the maximum amount of effective background improvement we can achieve by using SAC; the other 65\% of valid events are produced in Case B primary interactions that do not also produce a particle track in the LDA field (see Table \ref{tab:frame_summary}).

\begin{figure}[p]
\begin{center}
\includegraphics[width=4.0in]{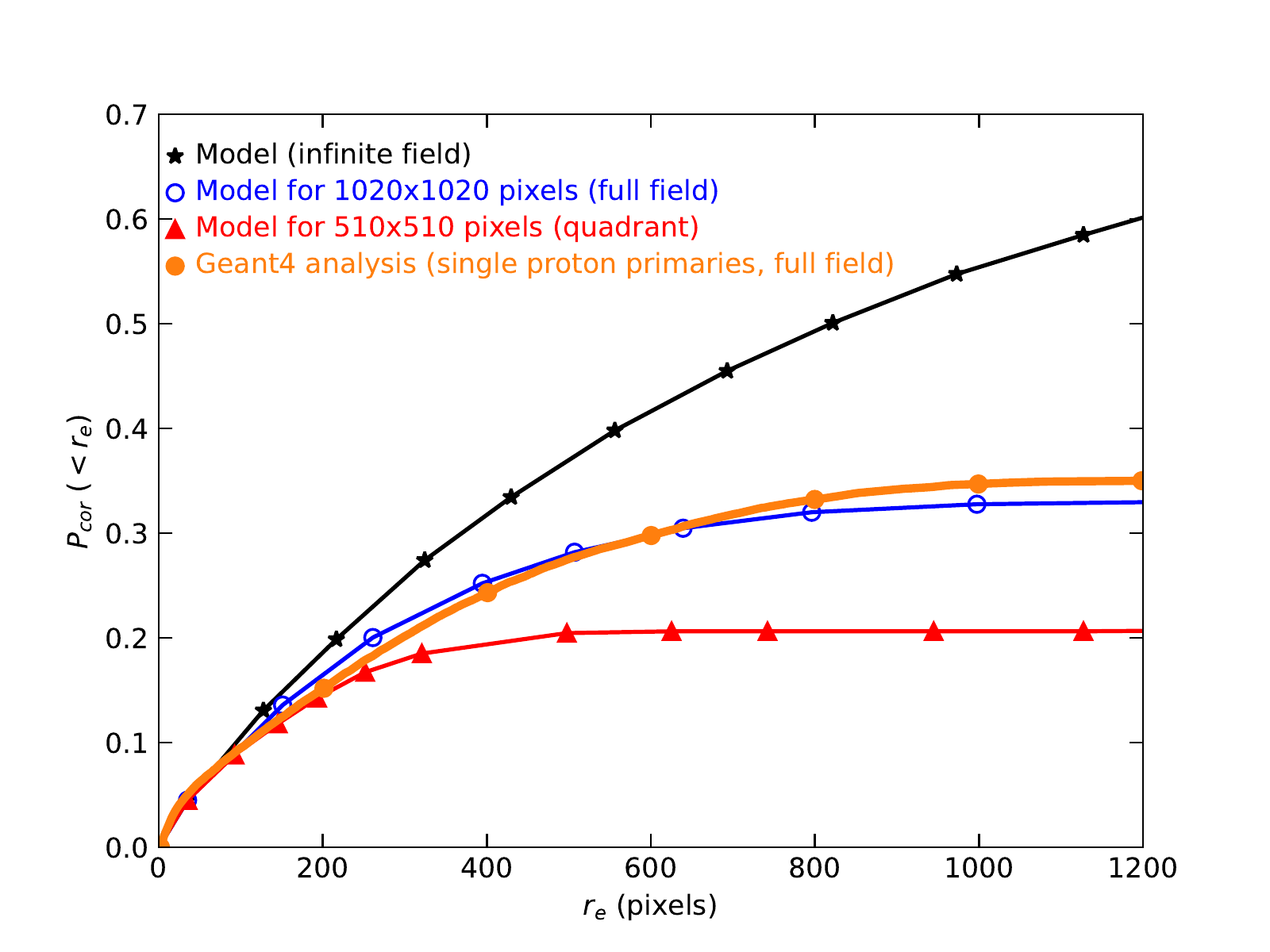}
\caption{Cumulative probability that a valid event falls within an exclusion radius $r_e$ of a particle track produced by the same primary. The orange line is derived from our single primary Geant4 simulation data. The other lines are from Figure \ref{fig:prob_cor} in Appendix \ref{sect:app_pcor}.}
\label{fig:pcor}
\end{center}
\end{figure}

\begin{figure}[p]
\begin{center}
\includegraphics[width=4.0in]{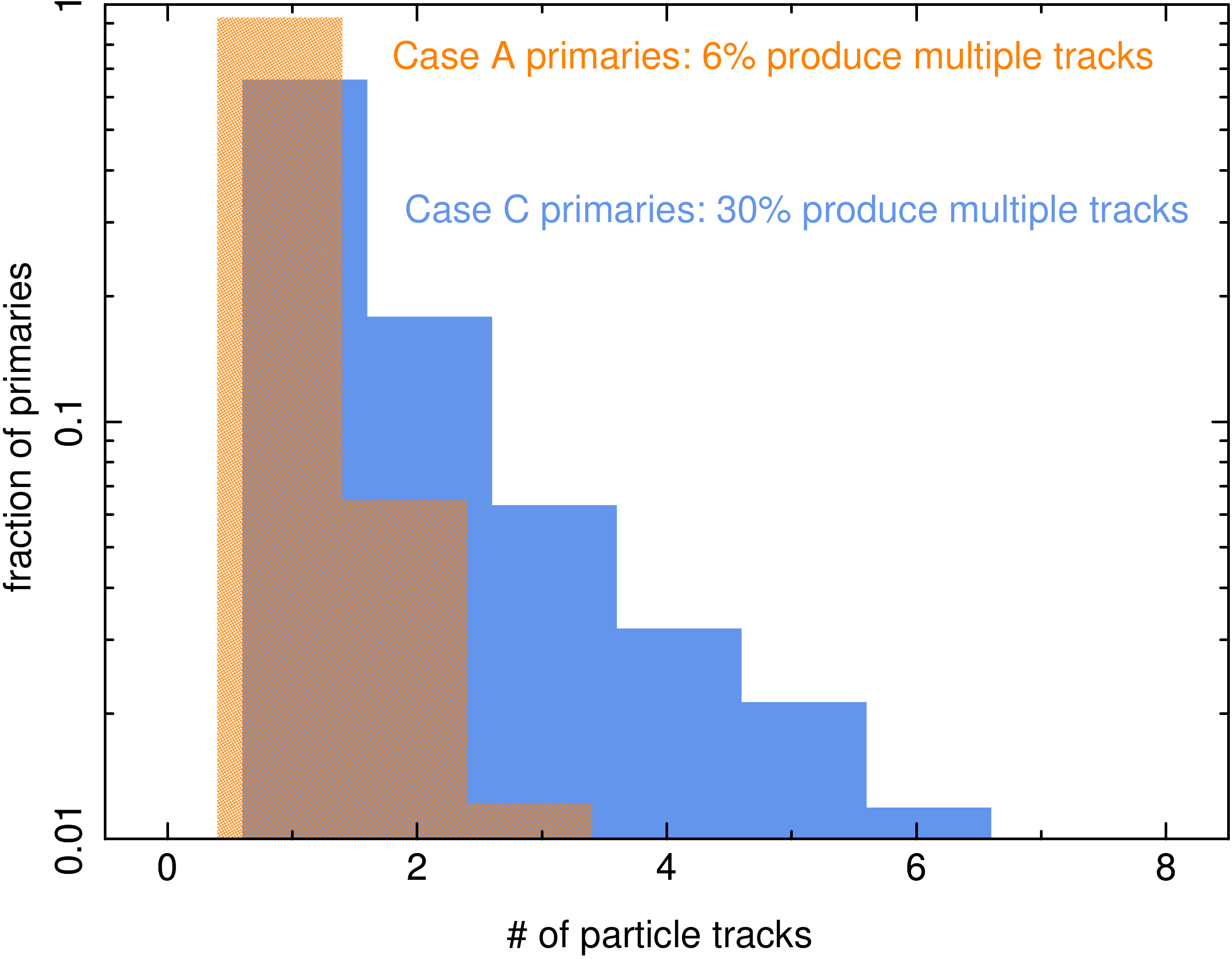}
\caption{Distribution of the number of particle tracks produced by primaries that do not also produce a valid event (Case A) and those that do (Case C). Valid events are more likely to be accompanied by a plurality of particle tracks. This can be used to identify frames that are likely to include valid events. Note that the histograms are shifted slightly along the X axis to improve clarity.}
\label{fig:havenot_numdist}
\end{center}
\end{figure}

In addition to a spatial correlation between particle tracks and valid events, we have found that proton primaries that produce valid events are much more likely to produce multiple particle tracks. This can also be seen from Table \ref{tab:frame_summary}, which shows that, among primaries that produce signal in the detector, Case A primaries produce on average 1.1 particle tracks, while Case C primaries produce 1.9 particle tracks. To further explore this, we plot in Figure \ref{fig:havenot_numdist} the distribution of particle track number for Case A and Case C primaries. Only 6\% of Case A primaries produce multiple particle tracks, whereas 30\% of Case C primaries do. Qualitatively, this makes sense; a primary interaction in the WFI structure can produce a shower of secondaries striking the detector, and these secondaries include both high-energy particles that produce tracks and lower energy photons and electrons that produce valid events. The number of independent particle tracks in a WFI frame contains some information about the likelihood of a valid event being present, and thus counting them could be a useful method to reduce the background. However, since this plurality occurs in 30\% of Case C primaries, and such primaries account for only 35\% of the valid events, no more than 10\% of the 2--7 keV background may be eliminated with this method. The potential gain is further reduced by the expectation of $\sim$3.5 particle tracks per 5-ms frame (see Table \ref{tab:frame_summary}). Nevertheless, we continue to explore ``multi-track'' selective SAC, whereby SAC is applied only on frames with a certain number of particle tracks, in the remainder of this work.

\subsection{Applying SAC to Geant4 Frame Data}
\label{sect:applying_sac}

We identify three metrics to represent improvement in the particle background. One is the simple level of the background which is used to define the WFI requirement. The other metrics, more relevant for certain Athena science cases, are the signal-to-background ratio, which is an estimator of systematic uncertainty; and the signal-to-noise ratio, an estimator of statistical uncertainty. Both are important in the background-dominated regime, although the level of importance depends on the details of the science goals being pursued. These metrics are derived in analytical terms in Appendix \ref{sect:app_sac}.

\subsubsection{Background Reduction and Lost Signal}
\label{sect:sandb}

The fractional reduced background is $b = B/B_o$, where $B_o$ is the original background before SAC is applied, and $B$ is the background after applying SAC masking, both measured in counts of valid events. Likewise, the fractional reduced signal is defined as $s$ = $S/S_o$, where $S_o$ is the original source signal (in counts) before SAC is applied, and $S$ is the signal after applying SAC masking. Unless noted otherwise, all of the metrics below using $s$ and $b$ are independent of the actual source or background flux, as shown in Appendix \ref{sect:app_sac}. In some cases this assumes that the observations are background dominated. We also assume the source is uniform spatially and temporally, so that $S/S_o$ goes as the fraction of area remaining after SAC, or $(1-A_R/A_T)$ in the notation of Appendix \ref{sect:app_sac}, where $A_R$ is the rejected area and $A_T$ is the total area.

\begin{figure}[t]
\begin{center}
\includegraphics[width=6.25in]{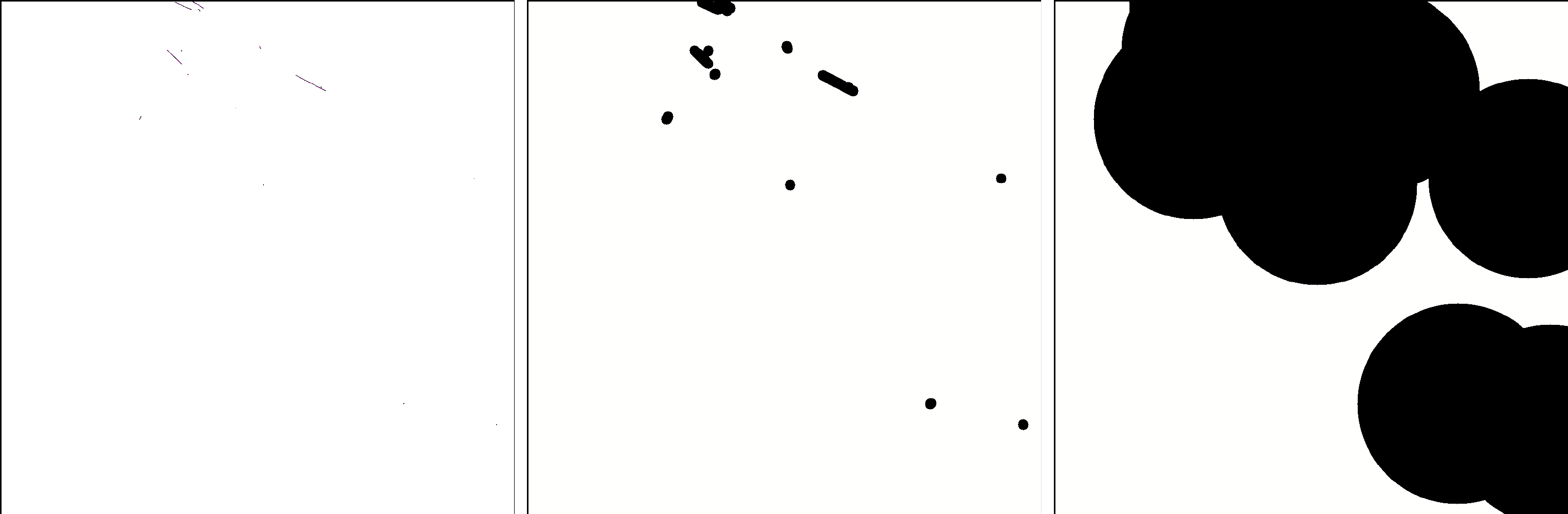}
\caption{
Example of calculating the fractional remaining area after SAC masking. The left image shows a single, full LDA frame with several particle tracks. The center image shows masking with a radius of 10 pixels around each particle track pixel, and the right frame shows masking with radius 200 pixels. Overlapping masking circles require us to empirically determine the remaining area.}
\label{fig:frame_mask_schematic}
\end{center}
\end{figure}

We calculate $s=S/S_o$ for a particular SAC exclusion radius $r_e$ directly from the Geant4-derived frame data. We create a mask for each frame and draw a circle of radius $r_e$ around each pixel in a particle track (or MIP pixel). Pixels inside these circles are set to zero, and those outside are set to one. The remaining fractional area $a_{good} = [1-A_R(r_e)/A_T]$, and thus $S(r_e)/S_o$, is simply the ratio of the amount of masked area to total area, summed over all frames. This method is shown schematically in Figure \ref{fig:frame_mask_schematic}. Unlike the Appendix \ref{sect:app_sac} method, which uses a Monte Carlo simulation to calculate this value, our method is subject to statistical limitation. For long frame time and large $r_e$, very little area is retained, so the uncertainty on $A_R$ becomes large.

We calculate $b=B/B_o$ for a particular SAC exclusion radius $r_e$ in a similar way from Geant4 results. In this case, we simply eliminate all valid events within $r_e$ of a MIP pixel, using the distance calculated in Section \ref{sect:event_finding}. $B(r_e)/B_o$ is then the number of remaining valid events divided by the original number. We note that $B/B_o$ is identical whether we restrict the valid events to 2--7 keV or use all valid events below 15 keV. Since the latter contains three times as many events as the former, we use all events below 15 keV to increase the statistics. Nevertheless, as for the masked area, for long frame time and large $r_e$ there are few valid events remaining and the uncertainty on $B/B_o$ becomes large. Where possible, we include these uncertainties in the following analysis.

\begin{figure}[t]
\begin{center}
\includegraphics[width=6.0in]{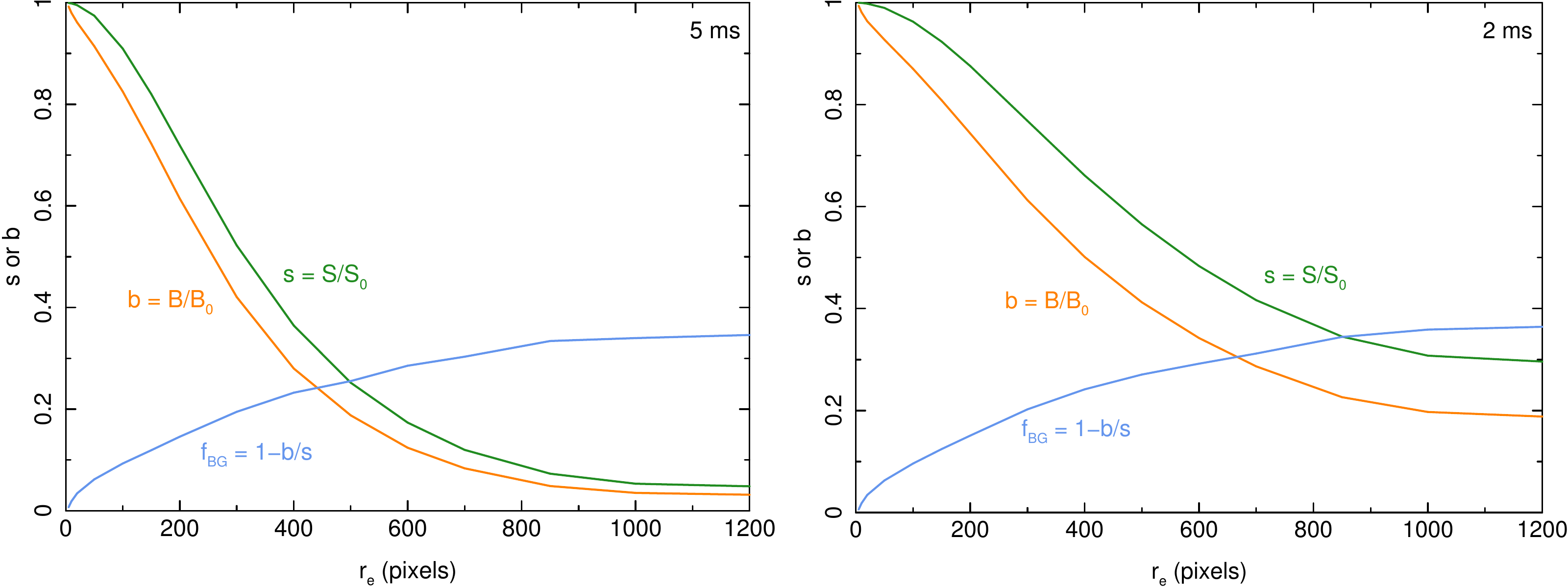}
\caption{Remaining signal $s$ and background $b$ as a function of SAC exclusion radius, for 5-ms (left) and 2-ms (right) frame times. The background reduction exceeds the signal loss at all masking radii. Also shown for reference is the fractional reduction in background surface brightness, $f_{BG}$, as a function of masking radius. This cannot exceed \pcor, shown in Figure \ref{fig:pcor}, since only correlated background events contribute to this improvement.}
\label{fig:sandb}
\end{center}
\end{figure}

In this notation, the first metric, the fractional reduction in background surface brightness $f_{BG}$, can be written as 
\begin{equation}
f_{BG}=(1-F_{BG}/F_{o,BG})\, ,
\end{equation}
where $F_{BG}$ and $F_{o,BG}$ are the reduced and original background surface brightness, respectively. Since 
\begin{equation}
F_{BG}=B/A_{good}
\end{equation}
and 
\begin{equation}
F_{o,BG}=B_o/A_T\, ,
\end{equation}
where the fractional remaining area is
\begin{equation}
a_{good} \equiv A_{good}/A_T = (1-A_R/A_T )=s\, ,
\end{equation}
some math tells us that 
\begin{equation}
f_{BG} = (1-b/s)\, .
\end{equation}
Since this is a surface brightness, it depends on both the remaining fractional area and the number of remaining background valid events. We plot this value as a function of $r_e$ in Figure \ref{fig:sandb}, along with $s$ and $b$. We further note that $f_{BG}$ cannot exceed \pcor\, as shown in Figure \ref{fig:pcor}, since only background events correlated with the masked particle track contribute to the background surface brightness reduction. Other background events are removed at a rate simply proportional to the lost area (denoted $P_{ran}$ in Section \ref{sect:app_sac_bkg}), so there is no reduction in surface brightness. However, this assumes the random background events are also uniformly distributed; as we showed in Figure \ref{fig:raddist}, this is not exactly the case.

\begin{figure}[t]
\begin{center}
\includegraphics[width=6.0in]{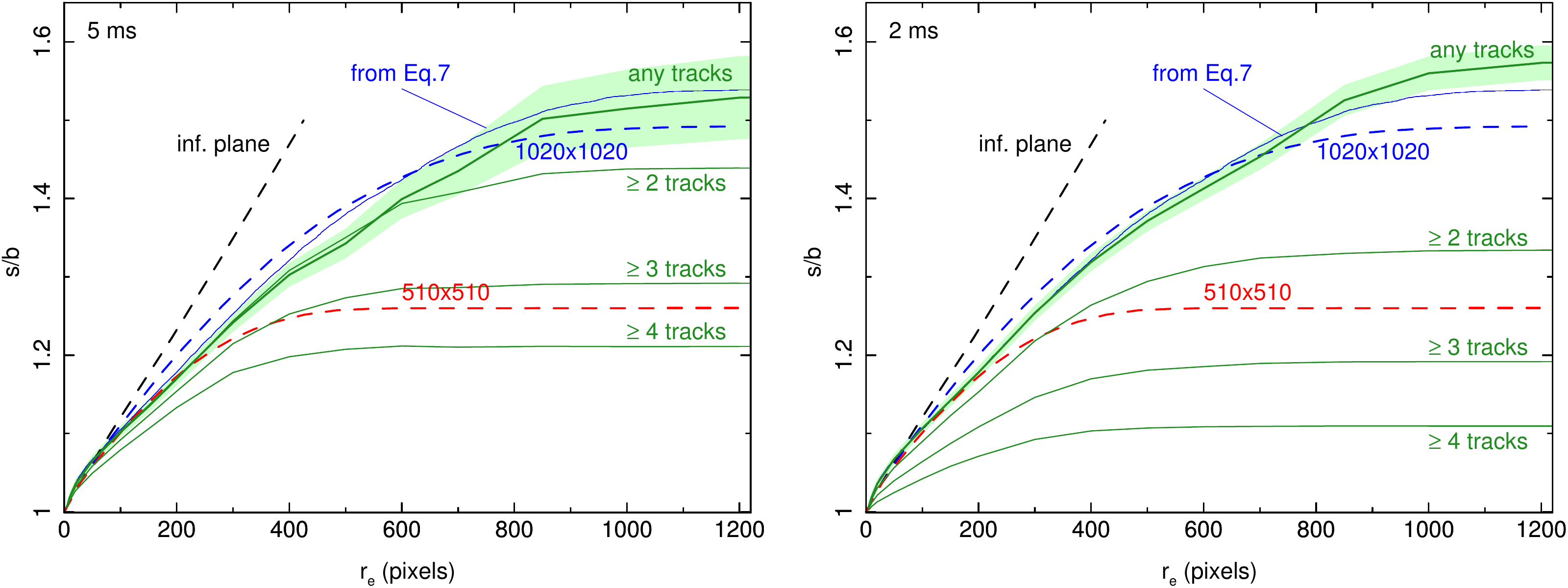}
\caption{Signal-to-background ratio for 5-ms (left) and 2-ms (right) frame time. Dashed lines correspond to lines of the same color in Figure \ref{fig:sb}, and are calculated from an analytic treatment of $s/b$ for an infinite plane (black), full LDA field (blue), and a single LDA quadrant (red) (see Appendix \ref{sect:app_sac_SB}). The solid blue line shows Eq.\ref{eq:sbrat} using $P_{cor}$ from Figure \ref{fig:pcor}, and is similar to the full-field analytic relation. The green curves show $s/b$ for different multi-track settings, with SAC enforced only on frames that contain at least the number of particle tracks shown. A 1\,$\sigma$ error region is shown for the ``any tracks'' curve for reference; this curve is measured directly from the simulation data, and the fact that it is largely consistent with both the dashed blue analytic curve and solid blue $P_{cor}$-derived curve provides an important cross-check of the methodology and different Geant4 simulations that informed each analysis.}
\label{fig:sbrat}
\end{center}
\end{figure}

\subsubsection{The Signal-to-Background Ratio}
\label{sect:sbrat}

The signal-to-background ratio, $s/b$, is an indicator of the systematic error in the measurement due to the irreducible limit of knowledge of the background. We plot this as a function of SAC exclusion radius in Figure \ref{fig:sbrat}, along with curves derived in Appendix \ref{sect:app_sac_SB} and Figure \ref{fig:sb}, for frame times of 5 and 2 ms.  The different green curves labelled ``any tracks'', ``$>$1 track'', etc., indicate the results from selective application of SAC only in frames that contain at least that many particle tracks. The ``any tracks'' curve corresponds to standard SAC, which masks around any particle track pixel in all frames. This curve is similar to the dashed blue curve calculated from
\begin{equation}
\label{eq:sbrat}
\frac{s}{b} = \frac{(S/S_o)}{(B/B_o)} =\frac{1}{(1 - P_{cor})} \, ,
\end{equation}
which is Eq.\ref{eq:sbn} derived in Appendix \ref{sect:app_sac_SB}. We calculate the empirical version of $s/b$ using the orange $P_{cor}$ curve in Figure \ref{fig:pcor} and show that as the thin blue line in Figure \ref{fig:sbrat}. This is consistent with the relation derived analytically in the Appendix, and also consistent with the directly determined ``any tracks'' curve, an important cross-check of the methodology and different Geant4 simulations that informed each analysis.

Since $s/b$ depends only on $P_{cor}$, there should be no change with frame time. The differences between 5 and 2 ms in the ``any tracks'' line (standard SAC) are due to statistical limitations in calculating the lost area and reduction in valid events from the background. The differences in multi-track selective SAC are real; for the shorter frame time, which is also a proxy for lower GCR flux, there are fewer particle tracks per frame, and thus fewer of the frames are participating in the background reduction.

It is clear that applying SAC to the full frame has a substantial benefit in this metric compared to applying it to a quadrant. This remains true when applying selective SAC only to frames with two or more particle tracks (``$>$ 1 track'' curve), at large exclusion radius.

\subsubsection{The Signal-to-Noise Ratio}
\label{sect:snr}

\begin{figure}[p]
\begin{center}
\includegraphics[width=6.25in]{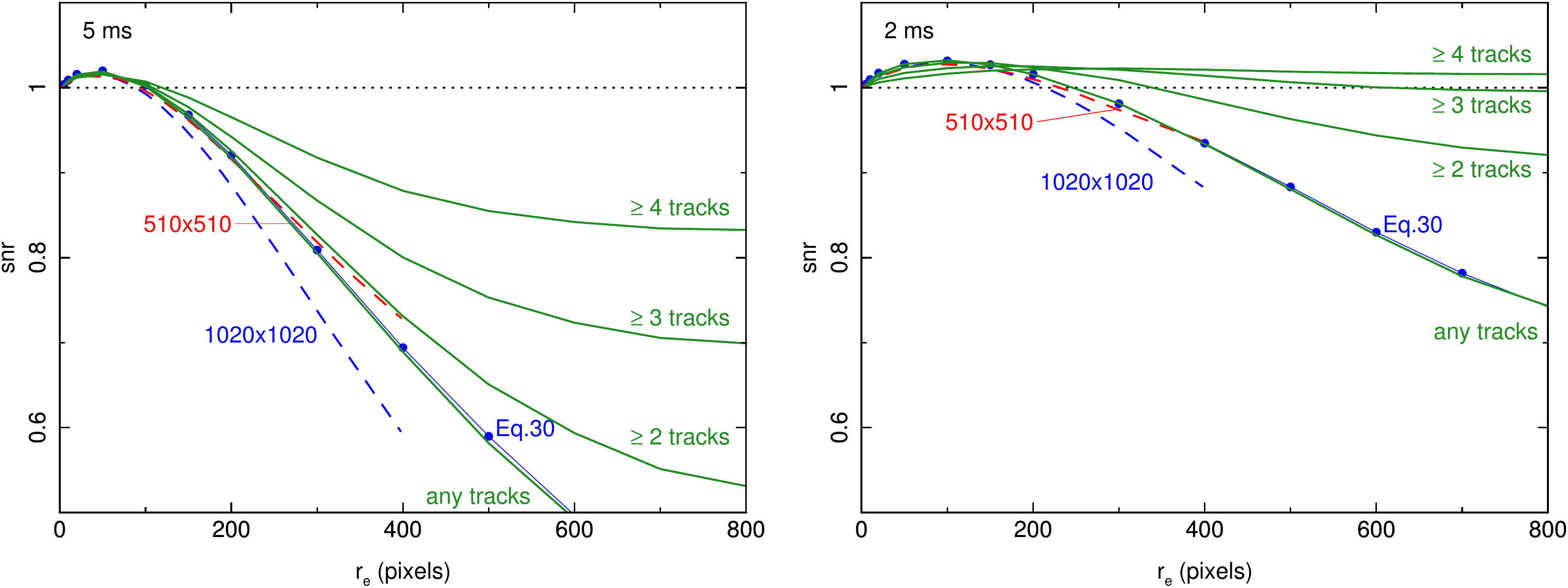}
\caption{Signal-to-noise ratio for 5-ms (left) and 2-ms (right) frame time. Dashed lines correspond to lines in Figure \ref{fig:app_snr} of Appendix~\ref{sect:app_sac_SNR}, and the blue line with circles shows our result of Equation \ref{eq:snrn} also from Appendix \ref{sect:app_sac_SNR}. Other notations are the same as in Figure \ref{fig:sbrat}. The multi-track method produces higher snr than standard SAC, especially at large exclusion radius. That the analytically derived relation is consistent with our empirically derived ``any tracks'' relation again provides a valuable cross-check for our methodology.}
\label{fig:snr}
\end{center}
\end{figure}

In the background-limited regime where $B \gg S$, and assuming counting statistics dominate any systematic errors, the signal-to-noise ratio is $SNR = (S/B)^{1/2}$. We follow Eq.\ref{eq:snr_sbsqrt} in Appendix \ref{sect:app_sac_SNR} and define a normalized signal-to-noise ratio as
\begin{equation}
{snr} \equiv \frac{S/S_o}{(B/B_o)^{1/2}} \, ,
\end{equation}
Our derived $snr$ curves are shown in Figure \ref{fig:snr} as a function of exclusion radius for 5 and 2 ms frame time, again with multiple curves for multi-track selective SAC, and including dashed lines derived in Appendix \ref{sect:app_sac_SNR}. The solid blue line shows the solution for Eq.\ref{eq:snrn} using our empirical $P_{cor}$, and this is fully consistent with the measured ``any track'' $snr$, indicating that the two independent Geant4 simulations produce compatible descriptions of $P_{cor}$. The derived $snr$ is similar to the dashed blue analytically derived curve, although they deviate at large exclusion radius. 

\begin{figure}[p]
\begin{center}
\includegraphics[width=6.25in]{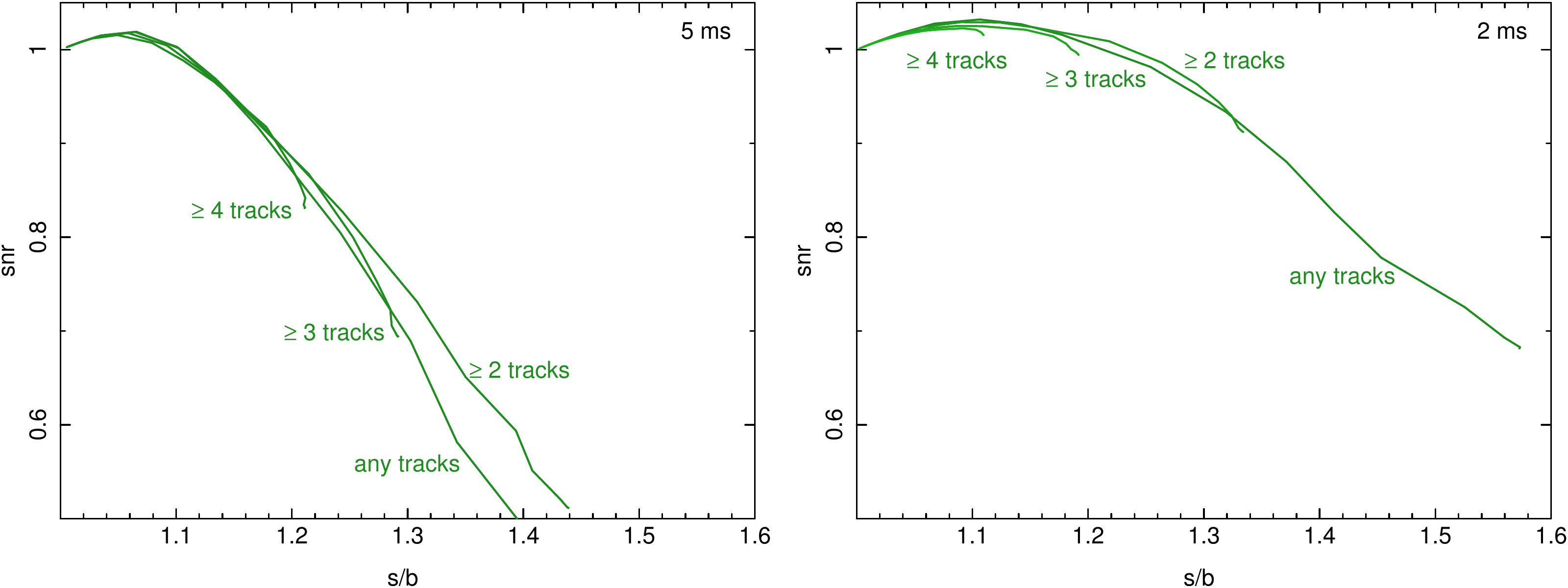}
\caption{Signal-to-noise ratio vs. signal-to-background ratio. Applying SAC in 5-ms frames with at least two particle tracks provides an improvement over standard SAC.}
\label{fig:snrvssbrat}
\end{center}
\end{figure}

Applying SAC to the full frame degrades the $snr$, especially at large exclusion radius, as more signal is lost. To explore the trade-off between $s/b$ and $snr$, in Figure \ref{fig:snrvssbrat} we plot $s/b$ vs. $snr$ for different settings of selective SAC. For 5-ms frame time, applying SAC in frames with at least two particle tracks provides a slight improvement over standard SAC.

\subsubsection{Effects of Rolling Shutter}
\label{sect:rollling}

We have made a simplifying assumption in the above that the full frame is read out instantaneously. In practice, the WFI LDA will implement a rolling shutter whereby each detector row is read out in sequence from top to bottom over the course of the 5-ms frame time. This means that any arriving cosmic ray may produce secondary particles that land on the other side of the current rolling shutter row, and since this happens virtually instantaneously compared to the speed of the rolling shutter, it results in the primary particle track and secondary events appearing in different frames. As we show analytically in Appendix \ref{sect:app_rolling_shutter}, this degrades the effectiveness of SAC, since it alters the spatial correlation between particle tracks and valid events by introducing a temporal dependence.

\begin{figure}[t]
\begin{center}
\includegraphics[width=6.5in]{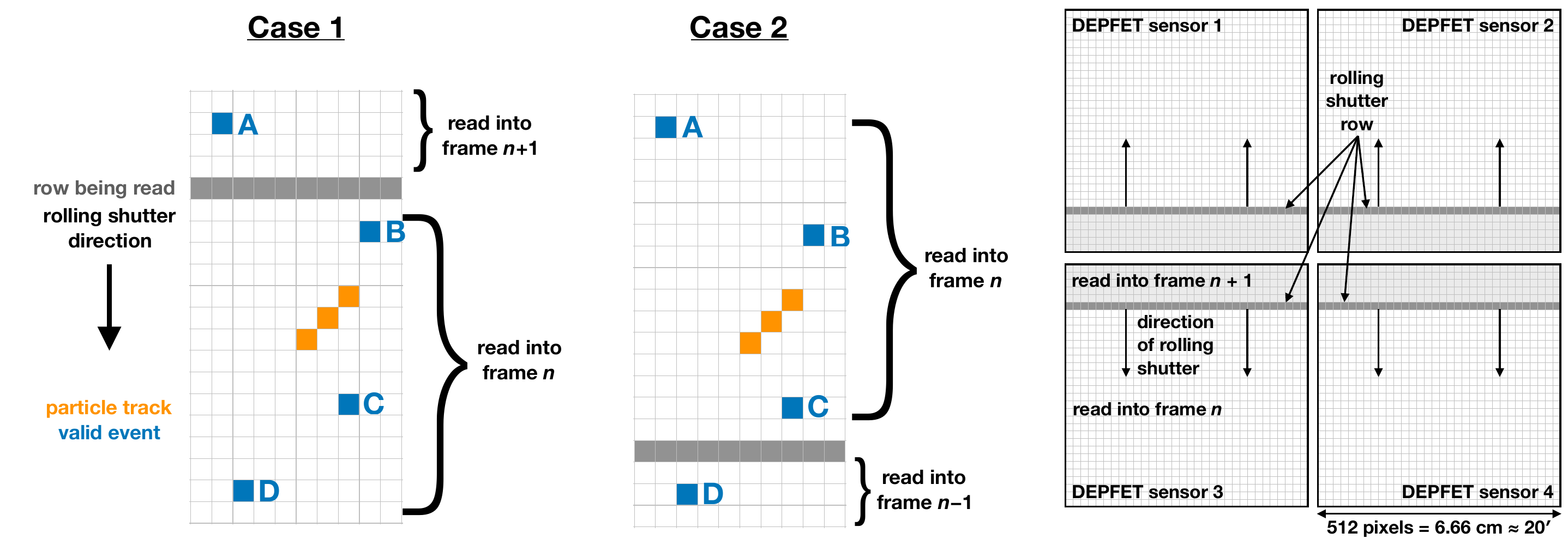}
\caption{Schematic showing the effects of the WFI LDA rolling shutter on SAC from the point of view of a MIP track that accompanies four valid events. The left two panels show the two major cases, where the MIP track is below (Case 1) or above (Case 2) the current readout row, which moves downward in these depictions. The right-most panel shows a depiction of the full LDA focal plane. We assume the rolling shutters are synchronized between the four DEPFET sensors and move in the directions shown.}
\label{fig:rolling_shutter_schematic}
\end{center}
\end{figure}

We approach this from the perspective of the particle track, since that is
a natural way for the SAC masking to be defined. We define frame $n$ as the
frame in which the particle track is recorded by the WFI, and we assume
that any pixels activated by the particle are done so instantaneously, in a
time much shorter than the row readout time ($\ll 10$ $\mu$s). This
includes pixels that are activated directly by the primary particle or by
any secondaries produced by interaction with the WFI structure. A schematic
of a particle interaction with a simplified WFI LDA is shown in Figure
\ref{fig:rolling_shutter_schematic}. The particle track of MIP pixels is
shown in orange. In this example, the particle produced four secondary
valid events, shown as blue pixels, which here cover all the possible
configurations of the particle track, valid events, and the current readout
row. There are two general cases: the current readout row is above the
particle track (Case 1), or it is below the particle track (Case 2), where
``above'' and ``below'' are defined for the rolling shutter moving downward. In Case 1, valid events B, C, and D are read into frame $n$ along with the particle track. Valid event A is above the rolling shutter, so it will be read into frame $n+1$.  In Case 2, valid events A, B, and C will be read into frame $n$ along with the particle track. Valid event D is below the rolling shutter and so is read into the frame currently being read, $n-1$. We ignore the case where a MIP track lands on the current readout row, which should occur for $<1$\% of MIP tracks. We finally assume that the rolling shutters on all four DEPFET sensors are synchronized, and that they operate as shown in the right panel of Figure \ref{fig:rolling_shutter_schematic}.

\begin{figure}[p]
\begin{center}
\includegraphics[width=4.0in]{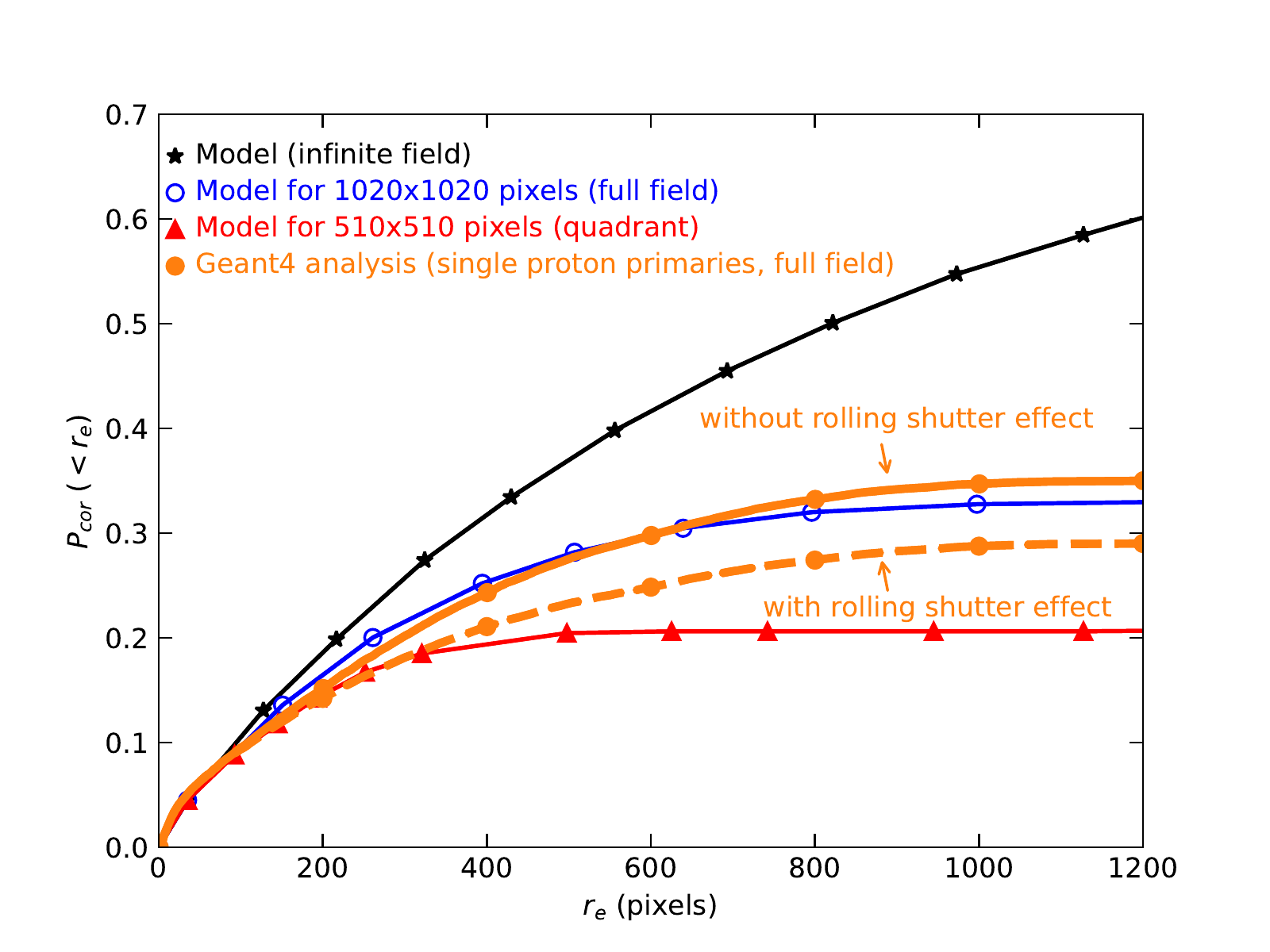}
\caption{Cumulative probability that a valid event falls within an exclusion radius $r_e$ of a particle track produced by the same primary, comparing the results without rolling shutter (from Figure \ref{fig:pcor}) and those including rolling shutter. There is very little difference at low $r_e$, since there is a low probability of the shutter row interloping between a particle track and its secondary events at these small distances. At larger $r_e$, the correlation degrades by about 20\%.}
\label{fig:pcor_rolling}
\end{center}
\end{figure}

We first determine the effects of rolling shutter on $P_{cor}$, the cumulative correlation between particle tracks and valid events they produce, using the single primary data set. For each primary, we randomly assign a rolling shutter row, and then for each particle track produced by that primary, we eliminate valid events on the other side of the shutter row before accumulating the probability distribution. The resulting $P_{cor}$ is shown in Figure \ref{fig:pcor_rolling} as a dashed orange line; comparison to the non-rolling-shutter $P_{cor}$ (solid orange line) shows little difference at small exclusion radius and a $\sim$20\% reduction at large exclusion radius. This makes qualitative sense. At the smallest particle track/valid event separations, it is very unlikely the rolling shutter will happen to fall between a particle track and its nearby secondary events. At intermediate separations, $r_e =$ 400--600 pixels, this becomes much more likely, and we see a large deviation of $\sim$ 20\% from $P_{cor}$ with no rolling shutter. At the  largest separations, there are very few particle track/valid event pairs contributing to the cumulative correlation, and so the rolling shutter effect is diluted and $P_{cor}$ remains about 20\% below the non-rolling-shutter value. This 20\% effect is less than the factor of two (or 50\%) estimated by the analytic treatment in Appendix \ref{sect:app_rolling_shutter}; the latter is really an upper limit, since it assumes the distribution of secondary valid events on the detector is random, rather than spatially correlated with the primary particle track as we have shown.

\begin{figure}[p]
\begin{center}
\includegraphics[width=6.25in]{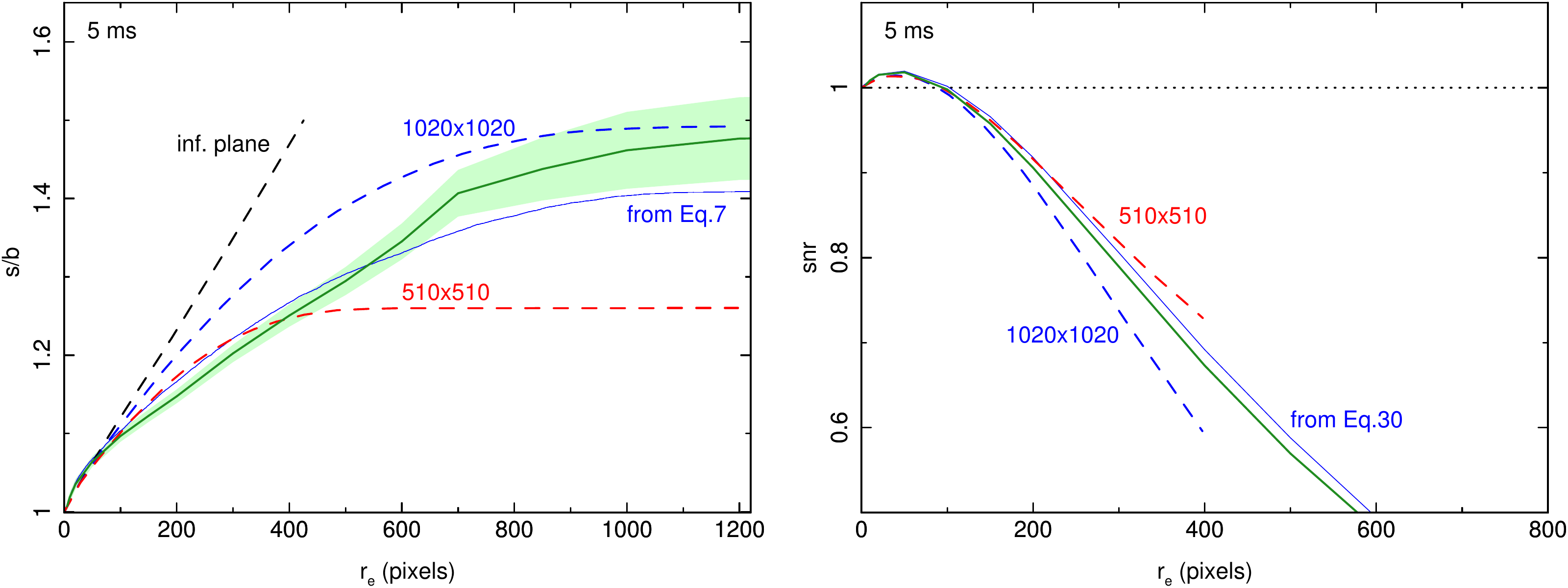}
\caption{(Left) Signal-to-background ratio for 5-ms frame time with rolling shutter included. Dashed lines are as in Figure \ref{fig:sbrat}. The solid blue line shows Eg.\ref{eq:sbrat} using $P_{cor}$ from Figure \ref{fig:pcor_rolling}, and demonstrates that the rolling shutter effect degrades this SAC background reduction metric at large exclusion radius ($s/b \approx 1.4$ with rolling shutter included compared to $s/b \approx 1.5$ without). The green curve and  1\,$\sigma$ error region is measured directly from the simulated rolling shutter data, and it is largely consistent with the solid blue $P_{cor}$-derived curve.
(Right) Signal-to-noise ratio for 5-ms frame time with rolling shutter included. Dashed lines are as in Figure \ref{fig:snr}. The thin blue line shows our result of Equation \ref{eq:snrn}, and is consistent with the green curve derived from rolling shutter simulations.}
\label{fig:sbrat_rolling}
\end{center}
\end{figure}

To quantitatively measure the effects of the rolling shutter on our SAC background reduction metrics, we adopt the ``minimal exclusion'' scheme described in Appendix \ref{sect:app_rolling_shutter}, whereby we only exclude valid events in the same recorded frame as a particle track, instead of also treating the preceding and trailing frames. Since we are including the effects of rolling shutter in the simulations but essentially ignoring them in the data analysis, this is a conservative approach to estimate the impact. We determine the signal-to-background ratio $s/b$ and signal-to-noise ratio $snr$ as described in Sections \ref{sect:sbrat} and \ref{sect:snr}, and show the results in Figure \ref{fig:sbrat_rolling}. We don't show results for the ``multi-track'' analysis here, but rather enforce SAC on frames with any number of MIP tracks. Once again, the empirically derived relations (green curves) are very similar to those calculated from the $P_{cor}$-based relations derived in Appendix \ref{sect:app_sac_SB} and \ref{sect:app_sac_SNR} (blue curves). The $s/b$ relation departs from what is shown in Figure \ref{fig:sbrat} without rolling shutter; the improvement in this metric at large exclusion radius is about 25\% lower with rolling shutter included in the simulated observations and the ``minimal exclusion'' SAC analysis scheme implemented. This is fully driven by the difference in $P_{cor}$. The $snr$ is not greatly different from the non-rolling-shutter version, and in any event the improvement in $snr$ is restricted to small $r_e$, where the rolling shutter has minimal impact.

In the remaining analysis, unless otherwise noted, we focus on the simplified simulations that exclude rolling shutter.

\subsection{Practical Mitigation of the Background Using SAC}
\label{sect:practical}

\begin{figure}[t]
\begin{center}
\includegraphics[width=6.0in]{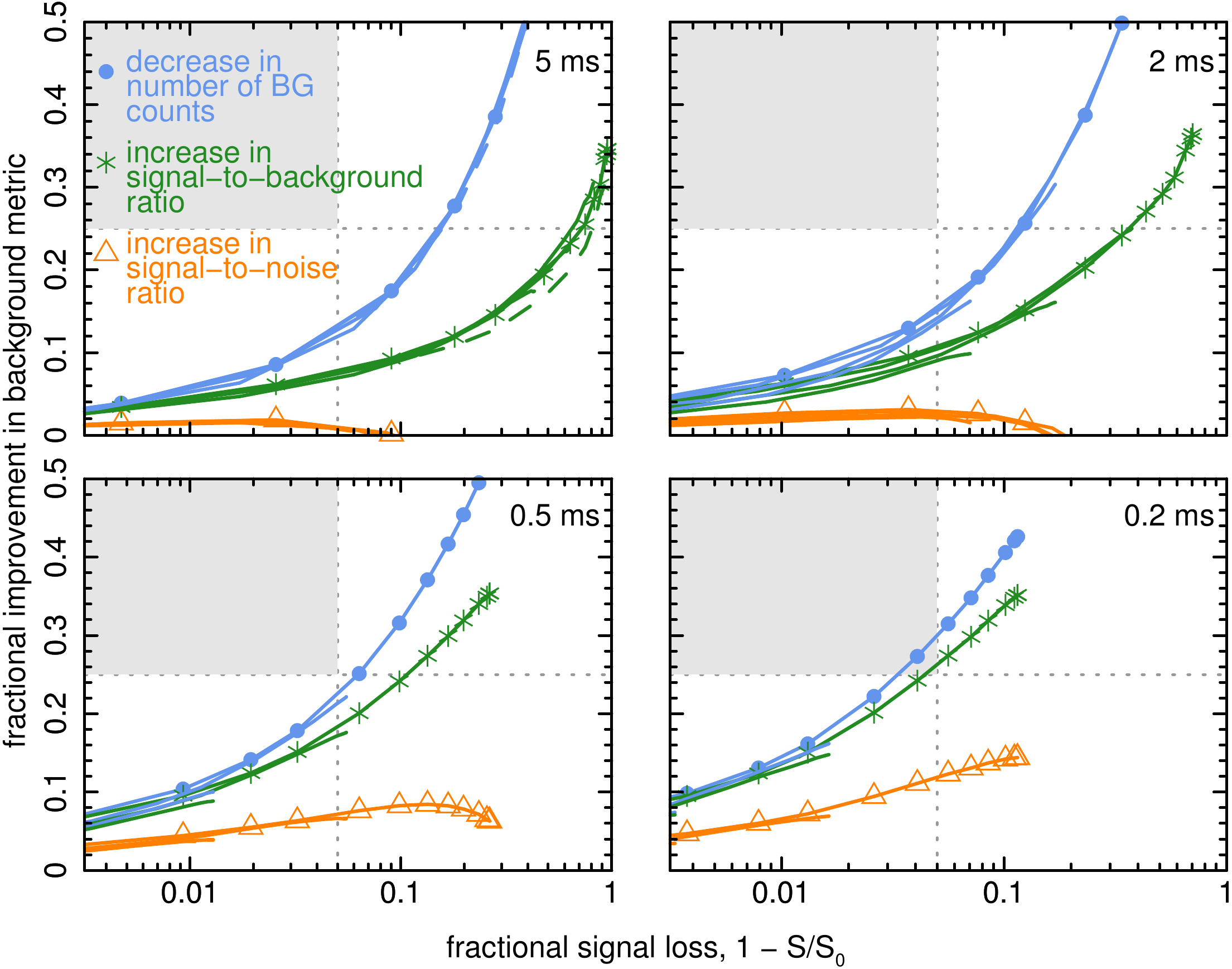}
\caption{Background ``improvement'' as a function of signal loss plotted for three different background reduction metrics, with different panels showing results for different frame times. Results from rolling shutter simulations using the ``minimal exclusion'' SAC scheme are shown by dashed curves in the 5-ms panel. The shaded quadrants show notional but arbitrary requirements that might be imposed for a background reduction technique: at least a 25\% improvement in the background metric (above the horizontal dashed line) accompanied by no more than 5\% signal loss (left of the vertical dashed line). The SAC technique cannot reach such requirements for any metric except in the shortest frame times.}
\label{fig:bgmetrics}
\end{center}
\end{figure}

The preceding analysis shows that, by employing SAC, we are able to reduce the background as measured by any of these three metrics:
\begin{itemize}
    \item Number of unrejected (valid) background events, $b = B/B_0$
    \item Signal-to-background ratio, $s/b = (S/S_0)/(B/B_0)$
    \item Signal-to-noise ratio, $snr = (S/S_0)/(B/B_0)^{1/2}$.
\end{itemize}
Here again, $B_0$ and $S_0$ are original background and signal counts, and $B$ and $S$ are background and signal counts that remain after SAC masking. Regardless of metric, the SAC background reduction is always accompanied by a loss of signal at the combination of expected frame rate and GCR flux for the WFI.  We show this in Figure \ref{fig:bgmetrics}, which plots the three metrics against the fractional reduction in signal counts. We also impose notional but somewhat arbitrary requirements on the SAC technique: it must improve the background by at least 25\% while reducing the signal by no more than 5\%. Here ``improvement'' depends on the metric; it can be a fractional reduction in the background count rate, or a fractional increase in $s/b$ or $snr$. The upper left quadrant satisfies these requirements, and at no point for either 5 ms or 2 ms frame time, or for enforcing different multi-track SAC thresholds, does the line pass through this quadrant. Only with short frame times of $<$0.5 ms do any of the metrics pass through the necessary quadrant. For these frame times, SAC remains useful, since the exclusion radius is smaller than the LDA field size. For very short times, normal anti-coincidence can be used, and the full frame excluded when a MIP pixel is detected.

On first glance, a simple reduction of background counts would appear to afford the best performance, but it is a specious metric since removal of relevant secondary background events is always accompanied by a similar removal of signal. Putting this metric aside,  we see that in all instances the improvement in $s/b$ is greater than that in $snr$. This reveals that SAC is more effective in reducing systematic errors than statistical ones. Indeed, for large fractional signal losses, improvement on $s/b$ is reached at the expense of a loss on $snr$. Even though we cannot meet the notional requirements, any improvement in $s/b$ can be useful as long as $snr$ does not suffer, and SAC can be thought of as a way of turning irreducible systematic errors into statistical errors that are reducible via an increase in exposure time. The analysis in the previous sections assumes that uncertainties are all statistical in nature, and that the noise term in $snr$ in the background-dominated regime is simply the square root of the number of valid events. Experience with deep observations of low surface brightness emission in XMM-Newton (e.g., cluster outskirts and galaxy halos) has shown that the limiting factor in these observations is never statistical uncertainty but always systematic uncertainty in the level of the background. For XMM-Newton EPIC pn observations, there is typically a 5\% irreducible uncertainty in the background that dominates the detection and characterization of faint diffuse emission; this is largely driven by the residual soft proton background\cite{Ghirardinietal2018}, which should be avoided on Athena through orbit selection and use of a magnetic diverter. However, for investigations of such low surface brightness sources, any reduction in the absolute level of the background via SAC could significantly improve the scientific return even if a significant number of source photons were discarded. 

Systematic uncertainty can arise from a number of sources depending on the strategy of the observations. For field-filling diffuse sources, often a non-contemporaneous blank-sky pointing is used to constrain both the focused X-ray and unfocused particle background, introducing systematic effects due to background time variability and changes in instrumental performance or calibration. Although quantifying these effects is complicated, in a simple model we can treat systematic uncertainty as a variance that adds linearly in the error budget, rather than in quadrature, and is thus not reducible by increasing the exposure time.  The SNR of a diffuse source observed by the WFI in such a case can be given by
\begin{equation}
\label{eq:diffsnr}
{SNR} = \frac{S_o}{(S_o+B_o+\sigma^2 B_o^2)^{1/2}} \,  ,
\end{equation}
where again $S_o$ is the number of source counts and $B_o$ is the number of background counts, where counts refer to valid events. These counts are related to the source and background surface brightness, $S_o^\prime$ and $B_o^\prime$, both in units of cts s$^{-1}$ arcmin$^{-2}$ integrated over some energy band by the relations
\begin{equation}
S_o = S_o^\prime \Omega t_{obs} \, ,
B_o = B_o^\prime \Omega t_{obs} \, ,
\end{equation}
where $\Omega$ is the solid angle of the region and $t_{obs}$ is the observing time. The value of $\sigma$ defines the systematic uncertainty expressed as a fraction of the background level, with $\sigma$ = 0.05 for a typical deep XMM-Newton observation\cite{Ghirardinietal2018} and $\sigma = 0.02$ as a current best estimate for Athena WFI, based on the requirement for knowledge of the non-focused particle background above 1 keV\cite{AthenaSciReq2.0.1}. For observations which are both background-dominated ($B_o \gg S_o$) and of small regions or short exposure times ($B_o \ll 1/\sigma^2$), Eq.~\ref{eq:diffsnr} reduces to the standard $SNR = S_o/B_o^{1/2}$. However, in the case where the systematic error of the background begins to dominate, $B_o \gg 1/\sigma^2$ and $SNR = S_o/(\sigma B_o)$. Increasing the exposure time in this case does nothing to increase the sensitivity because the uncertainty is dominated by uncertainty in the background level. This is the idea behind SAC; we remove background at the cost of signal, because that lost signal can always be recovered by increasing the exposure time.

\begin{figure}[p]
\begin{center}
\includegraphics[width=4.0in]{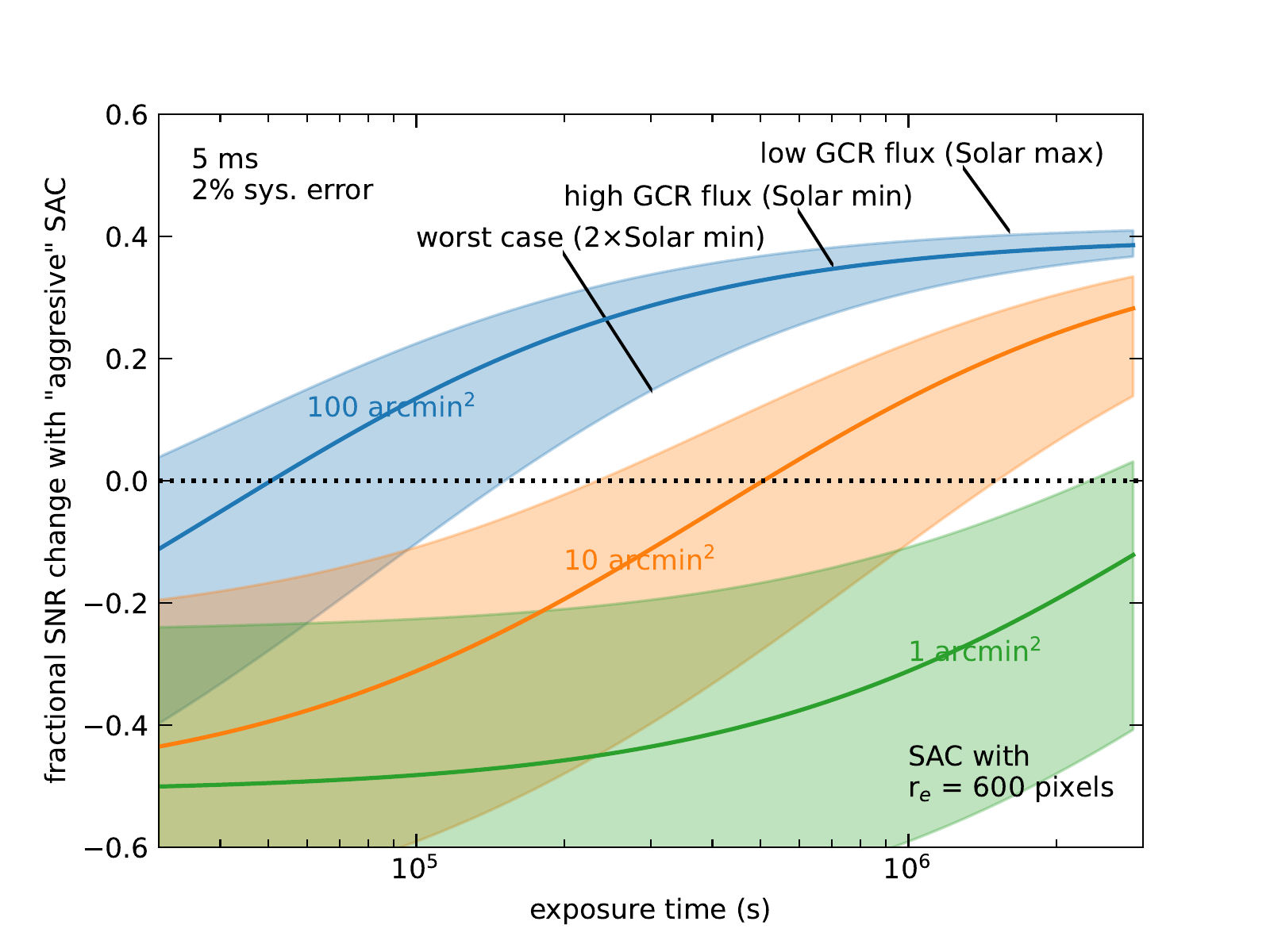}
\caption{Fractional change in SNR as a function of exposure time, comparing ``aggressive'' SAC with $r_e = 600$ pixels to not using SAC. This assumes a 2\% systematic error and a highly background-dominated observation. Shaded regions are shown for different sources sizes, and span the expected variation of the particle background; solid lines are from our best estimate of the maximum GCR flux at solar min, and lower limits are if these pre-launch estimates are low by a factor of two, similar to what is seen on eROSITA\cite{Freybergetal2020}. Even in this worst case scenario, SAC is still valuable at improving the SNR for the deepest exposures of the most extended faint sources.}
\label{fig:snrvstexp}
\end{center}
\end{figure}

Although loss of signal is usually undesirable, for some important WFI observations such as deep surveys, SAC can provide significant improvements in surface brightness sensitivity that yield important science. An example is shown in Figure \ref{fig:snrvstexp}, where we show the change in SNR as a function of exposure time for a number of source sizes by applying SAC aggressively with exclusion radius $r_e = 600$ pixels. We use Eq.\ref{eq:diffsnr} to calculate SNR in the presence of systematic error, using $\sigma = 0.02$ (2\%) as our current best estimate from the Athena WFI particle background knowledge requirement\cite{AthenaSciReq2.0.1}. Since we plot fractional change in SNR compared to not using SAC, the source flux cancels out in the assumed highly background-dominated limit (see also the derivation in Appendix \ref{sect:app_sac_SNR}). Shaded regions show results for sources of of 1, 10, and 100 arcmin$^2$ in extent, sampling typical sizes of low-surface-brightness features that might yield interesting science. The shading spans the expected variation of the particle background over an extended Athena mission: solid lines are from our best estimate of the maximum GCR flux at solar min, upper limits correspond to predicted minimum GCR flux at solar max, and lower limits are if the pre-launch estimates are low by a factor of two, similar to what is seen on eROSITA\cite{Freybergetal2020}. SAC offers a substantial improvement on large scales regardless of exposure time, and still significant improvement on smaller scales, especially during times of low background.

\begin{figure}[p]
\begin{center}
\includegraphics[width=6.25in]{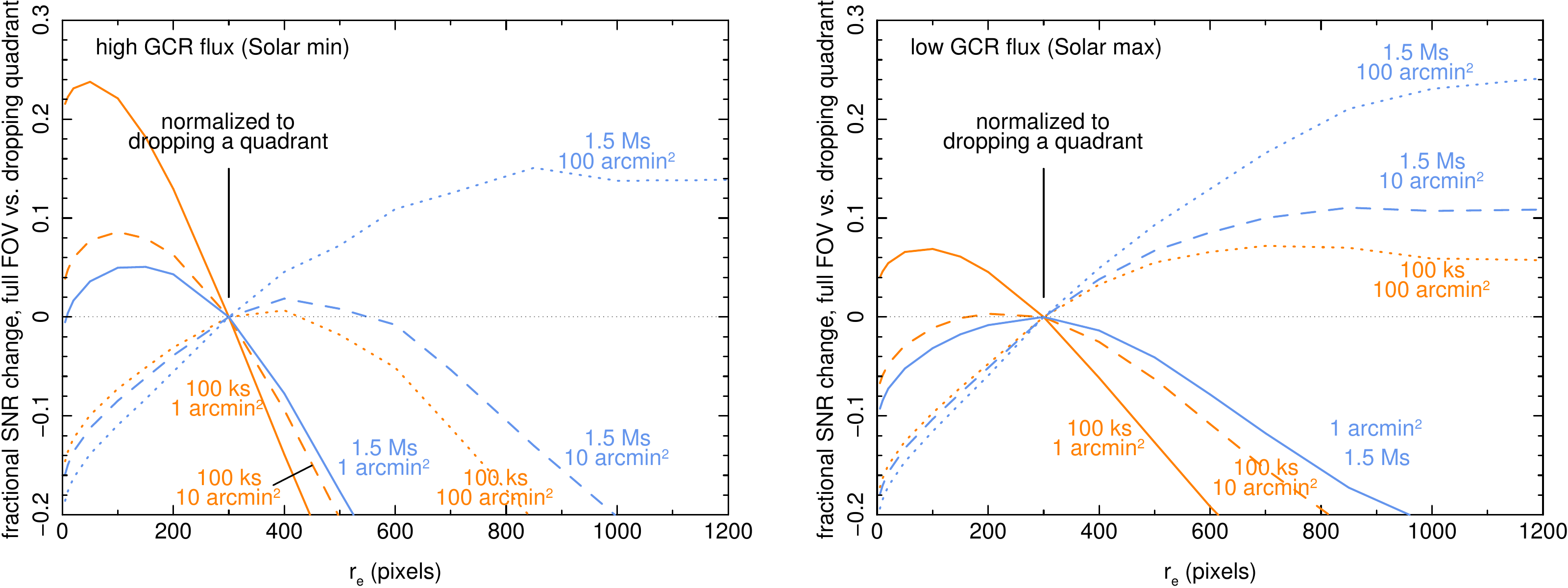}
\caption{Fractional change in SNR by SAC as a function of exclusion radius for times of high (left) and low (right) GCR flux expected over the life of an extended Athena mission. Here ``change'' is compared to the SNR achieved by eliminating a full quadrant, simulated here as an exclusion radius of 300 pixels. A value of zero means there is no SNR change. The model, described in the text, is a diffuse, faint emission source observed for 100 ks (orange) and 1.5 Ms (blue), and searched for extended features of 1 (solid), 10 (dashed), and 100 (dotted) arcmin$^2$ in size. This includes a systematic background uncertainty of 2\%. SAC over the full FOV provides flexible, selective masking out to large exclusion radius, inaccessible to single-quadrant SAC, that greatly enhances deep exposures of extended sources.}
\label{fig:snrimp}
\end{center}
\end{figure}

The ``selective'' SAC masking described here works best if the full LDA field is utilized, rather than a single quadrant. The optimal radius to mask depends sensitively on the exposure time and the size of the region, which both depend on the science under study. For an illustration of the power of SAC, we simulated a diffuse source of 2 keV thermal emission matching the Athena WFI surface brightness requirement of $6.2\times10^{-16}$ erg cm$^{-2}$ s$^{-1}$ arcmin$^{-2}$ in the 5--7 keV band\cite{AthenaSciReq2.6}. We simulated two levels of accompanying particle background surface brightness, one at $5.5\times10^{-3}$ cts cm$^{-2}$ s$^{-1}$ keV$^{-1}$ in the 2--7 keV band, from the WFI requirements to represent solar min (left panel), and one at half this flux to represent solar max (right panel). From this we estimated source and background counts in observations of 100 ks and 1.5 Ms, bracketing the depths of the WFI shallow and deep survey, and in regions of 1, 10, and 100 arcmin$^2$, as in Figure \ref{fig:snrvstexp}. Using Eq.~\ref{eq:diffsnr} and our empirical relations of signal loss $s$ and background loss $b$ (see Figure \ref{fig:sandb}), we then determined the SNR as a function of SAC exclusion radius $r_e$, assuming a systematic background uncertainty of 2\% ($\sigma = 0.02$), the Athena WFI requirement. We plot this in Figure \ref{fig:snrimp} as the fractional SNR change over the maximum background reduction possible with single-quadrant SAC, equivalent to dropping a quadrant containing a MIP pixel. Here we simulate this as the SNR at $r_e = 300$ pixels, where $s/b$ for single-quadrant SAC starts to turn over and become less effective (red line in Figure \ref{fig:sbrat}).

For shorter exposures sampling small regions, there is little improvement over single-quadrant SAC. However, for the deepest WFI exposures, and for science that requires detecting coherent structures on large scales, applying SAC to the full FOV can improve the SNR by up to $\sim$20\% over single-quadrant SAC, with no increase in exposure time. This is especially true for times of lower GCR flux. To obtain this improvement with single-quadrant SAC would require an additional 600 ks observation of this field. 

It is of course clear that many science cases will not benefit from masking regions around particle tracks. But allowing the application of SAC by the science observer maximizes the science return in a way that simply dropping quadrants or full frames would not. By telemetering every MIP pixel location in each frame, or at the very least the distance to the closest MIP pixel for each in-band event, WFI data would enable flexible application of SAC masking. This method has no impact on science investigations that do not benefit from it, as the science observer could decide whether to use it or not, or even experiment with different values of masking radius.

%%%%%%%%%%%%%%%%%%%%%%%%%%%%%%%%%%%%%%%%%%%%%%%%%%%%%%%%%%%%%%%
% Summary
%%%%%%%%%%%%%%%%%%%%%%%%%%%%%%%%%%%%%%%%%%%%%%%%%%%%%%%%%%%%%%%
\section{Summary}
\label{sect:summary}

We have presented an analysis of Geant4 simulations of the Athena WFI particle background in an effort to mitigate its effects. The majority of simulated 5-ms frames (87\%) contain only particle tracks that cannot be confused with focused X-rays due to their morphology or total energy; an additional 8\% of frames contain both particle tracks and X-ray-like events. This means that true anti-coincidence techniques cannot be used to drop frames, as it would remove $\sim$95\% of the source signal. We have developed and presented a partial veto scheme called Self-Anti-Coincidence, or SAC, which exploits a spatial correlation between particle tracks and secondary valid events, a correlation that we have identified and validated with independent Geant4 simulations and in-flight XMM-Newton EPIC pn data. By masking smaller regions of the FOV around particle tracks, this technique can greatly reduce the systematic effects of particle background in certain science cases, most notably observations of very faint, highly extended sources. With sufficient information included in WFI telemetry, this filtering can be applied selectively on the ground by the user, enabling detection of very low surface brightness objects without sacrificing other science.

We stress that the work presented here is not restricted to the Athena WFI, but is relevant for any future silicon-based pixelated X-ray imaging detector. In addition to providing a novel background mitigation technique for the WFI, the results and methodology can be used to generate requirements on elements such as frame rate, detector size, and particle environment for future missions. Such considerations will maximize the science return from otherwise challenging observations of very faint, extended X-ray sources.

%%%%%%%%%%%%%%%%%%%%%%%%%%%%%%%%%%%%%%%%%%%%%%%%%%%%%%%%%%%%%%%
% Acknowledgements
%%%%%%%%%%%%%%%%%%%%%%%%%%%%%%%%%%%%%%%%%%%%%%%%%%%%%%%%%%%%%%%
\section{Acknowledgements}
\label{sect:ack}

This work was done under the auspices of the Athena WFI Background Working Group, a consortium including MPE, INAF/IASF-Milano, IAAT, Open University, MIT, SAO, and Stanford. We thank the entire Working Group for valuable discussions that contributed greatly to this paper. We also thank the anonymous referee for helpful comments that significantly improved the manuscript. The US-based co-authors gratefully acknowledge support from NASA grant NNX17AB07G. The studies undertaken at the Open University were funded by the UK Space Agency, for which the team are grateful for their ongoing support.

This paper made use of simulations from Geant4 software\cite{Geant4,Geant4_paper2} and utilized the following software libraries for data analysis and presentation:
the Perl Data Language (PDL, \url{pdl.perl.org}) developed by K. Glazebrook, J. Brinchmann, J. Cerney, C. DeForest, D. Hunt, T. Jenness, T. Lukka, R. Schwebel, and C. Soeller;
NumPy\cite{NumPy} (\url{numpy.org}); 
Astropy\cite{astropy:2013, astropy:2018} (\url{http://www.astropy.org}), a community-developed core Python package for Astronomy;
and Matplotlib\cite{Matplotlib} (\url{https://matplotlib.org}), a Python library for publication quality graphics.

%%%%%%%%%%%%%%%%%%%%%%%%%%%%%%%%%%%%%%%%%%%%%%%%%%%%%%%%%%%%%%%
% References
%%%%%%%%%%%%%%%%%%%%%%%%%%%%%%%%%%%%%%%%%%%%%%%%%%%%%%%%%%%%%%%
%\bibliography{edm}   % bibliography data in report.bib
\bibliographystyle{spiejour}   % makes bibtex use spiejour.bst

%%%%%%%%%%%%%%%%%%%%%%%%%%%%%%%%%%%%%%%%%%%%%%%%%%%%%%%%%%%%%%%
% Biographies of authors
%%%%%%%%%%%%%%%%%%%%%%%%%%%%%%%%%%%%%%%%%%%%%%%%%%%%%%%%%%%%%%%
\vspace{2ex}\noindent\textbf{Eric D.~Miller} is a research scientist at the MIT Kavli Institute for Astrophysics and Space Research. He received his BA degree in physics from Oberlin College in 1996, and his MS and PhD degrees in astronomy and astrophysics from the University of Michigan in 1998 and 2003, respectively. He leads the XRISM In-Flight Calibration Planning Team, develops X-ray imaging detectors for future missions, and studies galaxy clusters and the diffuse intergalactic medium.

\vspace{1ex}
\noindent Biographies and photographs of the other authors are not available.

%%%%%%%%%%%%%%%%%%%%%%%%%%%%%%%%%%%%%%%%%%%%%%%%%%%%%%%%%%%%%%%
% Appendix 1
%%%%%%%%%%%%%%%%%%%%%%%%%%%%%%%%%%%%%%%%%%%%%%%%%%%%%%%%%%%%%%%
\appendix

\section{Self-Anti-Coincidence (SAC) Estimators}
\label{sect:app_sac}
% the background, $B$; the signal, $S$; the signal to background ratio, $S/B$ and the signal to noise ratio $SNR$
In this Appendix we introduce several quantities which may be used to assess the effectiveness of the SAC technique and explore their relationship with key parameters such as the frame time, $t_f$, and the exclusion radius, $r_{e}$.

\subsection{Signal}
\label{sect:app_sac_signal}
Let us start with the signal, $S$. We define $S$ as the source valid event counts accumulated over the region of interest, such as an LDA quadrant or the full LDA, per unit frame, averaged over many frames. Here we shall assume the signal to be distributed uniformly over the region of interest. The probability that, in a given frame, a valid event is lost due to SAC is $P_{sl} \equiv l_s/t_s$, where $l_s$ is the number of lost events and $t_s$ is the total number of events. Under the assumption of spatial uniformity of the signal, we have:
\begin{equation}
\label{eq:p_sl}
P_{sl}= A_R /A_T \, ,
\end{equation}
where $A_R$ is the area masked or rejected by SAC and $A_T$ is the total area. From this we derive the expression for the signal:
\begin{equation}
\label{eq:s}
S = (1 - A_R /A_T) \cdot S_o \, ,
\end{equation}
where $S_o$ is the signal when no SAC is applied. In our calculations we will make use of the fractional signal loss $(S_o - S)/S_o$, which can be expressed as:
\begin{equation}
\label{eq:sloss}
(S_o - S)/S_o =  A_R /A_T\, .
\end{equation}
By comparing Eq.\ref{eq:p_sl} with Eq.\ref{eq:sloss} we see that the fractional signal loss and the rejection probability are actually the same thing:
\begin{equation}
P_{sl} = (S_o - S)/S_o \, . 
\end{equation} 
For small signal losses, i.e., $(S_o - S)/S_o \ll 1$, exclusion regions do not overlap and we can derive a simple formula explicitly relating the signal loss to the frame time and the exclusion radius. Indeed:
\begin{equation}
(S_o - S)/S_o = \frac{\pi r_e^2 \cdot N_p}{A_T}\, ,
\end{equation} 
where $\pi r_e^2$ is the area of a single exclusion region and $N_p$ is the number of particle tracks falling in a given frame. This assumes that the masking regions are circular and that particle tracks are small compared to the exclusion radius. By rewriting $N_p$ as the rate of cosmic ray particle tracks $cr_p$ over the region of interest (a quadrant or the full LDA) times the frame time $t_f$ we find:
\begin{equation}
\label{eq:fsloss}
(S_o - S)/S_o =  \frac{\pi r_e^2 \cdot cr_p \cdot t_f}{A_T}\, .
\end{equation}
 Eq.\ref{eq:fsloss} shows that the signal loss scales quadratically with the exclusion radius and linearly with the frame time. As already pointed out, this derivation is strictly correct in the linear regime, i.e., $(S_o - S)/S_o \ll 1$; as the signal loss increases, the probability that different exclusion circles overlap must be accounted for. A simple algebraic expression can also be worked out for very large exclusion radii. When the exclusion circles encompass the total area, the surviving signal will be associated with the fraction of frames in which no particle track appears on the detector. This fraction is $\exp(-N_p)$, assuming a Poisson distribution of arriving primaries. From this we derive:
\begin{equation}
\label{eq:fsl_sat}
(S_o - S)/S_o = 1 - \exp(-cr_p \cdot t_f) \, .
\end{equation}

Note that in the linear regime, the fractional signal loss does not depend upon the specific size of the region under consideration, quadrant or full FOV; indeed in Eq.\ref{eq:fsloss} the area dependence is found both in the numerator ($cr_p$) and in the denominator ($A_T$) and cancels out. Conversely, when we approach saturation, area does matter, as shown in Eq.\ref{eq:fsl_sat}, where the term in the exponent scales linearly with the total area through $cr_p$. This is quite intuitive:  the larger the area under consideration, the larger the number of particle tracks and the smaller the likelihood that, for a given frame time, a frame is track-free.

\begin{figure}[t]
\begin{center}
\includegraphics[width=\linewidth]{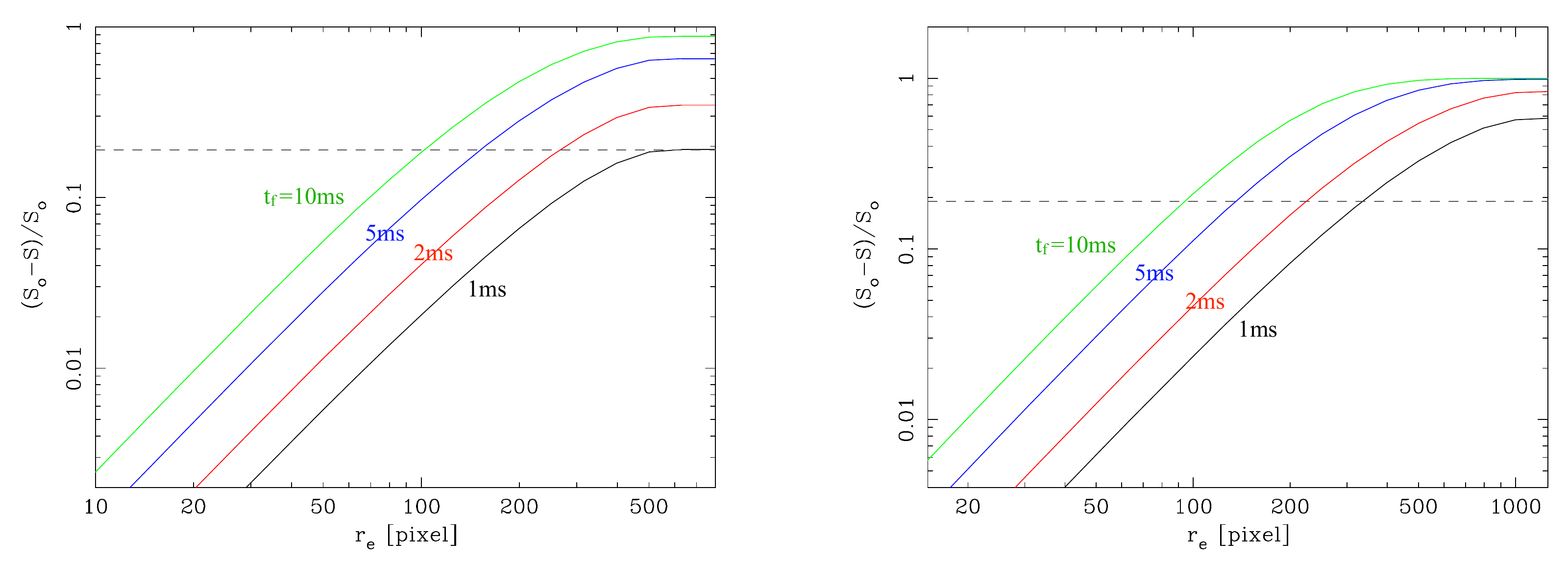}
\caption{Fractional signal loss as a function of exclusion radius for four different values of the frame time, as indicated by different colors. The left panel, for a 510$\times$510 grid, represents a single WFI LDA quadrant; the right panel, for a 1020$\times$1020 grid, represents the full LDA FOV.}
\label{fig:sig_loss}
\end{center}
\end{figure}

In the intermediate regime of signal loss between Eqs.\ref{eq:fsloss} and \ref{eq:fsl_sat}, masked areas overlap and there is no simple formula to estimate $(S_o - S)/S_o$. Thus we have resorted to Monte Carlo simulations. We assumed a primary frame-rate consistent with that reported in Ref.\ \citenum{vonKienlin2018} and a time resolution of 1 ms. We drew primary events in each bin following Poisson statistics and assigned random positions over 510$\times$510 and 1020$\times$1020 grids representing respectively a single LDA quadrant and the full LDA FOV. Finally, we re-binned the time-series to the desired frame time and computed the average rejected area, $A_R$, over a large number of frames. In Fig.\ref{fig:sig_loss} we plot the simulated fractional signal loss as a function of exclusion radius for four different values of the frame time: 1 ms, 2 ms, 5 ms, and 10 ms. For a given value of the frame time, signal loss increases with increasing $r_e$, quadratically in the linear regime (see Eq.\ref{eq:fsloss}) and saturating at large $r_e$ (see Eq.\ref{eq:fsl_sat}). As frame time goes down, the saturation regime shifts to larger exclusion radii. In other words, larger exclusion radii can be accepted for smaller frame times. In the limiting case of $t_f\rightarrow 0$ the exclusion circle can encompass the whole detector with no signal loss.

Let us focus on the left panel of Fig.\ref{fig:sig_loss}, which refers to a single quadrant, and assume we are willing to accept a certain fractional signal loss, say 20\%; we can distinguish three different regimes.
\begin{enumerate}
\item $t_f > 10$ ms: Fractional signal loss rapidly saturates to unity, only small exclusion radii can be accepted; in this regime SAC is of little or no use.
\item $t_f< 1$ ms: The exclusion circle can encompass the whole quadrant with an acceptable signal loss; this is the standard anti-coincidence regime.
\item 1 ms $<t_f<$ 10 ms: The exclusion circle is not restricted to very small values, however it cannot extend to the whole area. This in an intermediate regime where partial vetoing of the quadrant can be of use.
\end{enumerate}
Since the values of $t_f$ marking the transitions between the three regimes depend mostly on the value of the signal loss at saturation, analogous values for the full detector case depicted in the right panel of Fig.\ref{fig:sig_loss} can be obtained by dividing all frame times by a factor of four (see Eq.\ref{eq:fsl_sat}). Thus, for the full detector case we have:
\begin{enumerate}
\item $t_f > 2.5$ ms: SAC not practical;
\item $t_f< 0.25$ ms: standard anti-coincidence regime;
\item 0.25 ms $<t_f<$ 2.5 ms: partial vetoing regime.
\end{enumerate}
These cases are again for an acceptable signal loss of 20\%, and the acceptable level depends strongly on the particular sources and science under study. More importantly, the signal $S$ does not provide a full description of SAC, as it does not contain any information on the \textit{improvements} afforded by this partial vetoing technique. For this we must turn to other indicators.

\subsection{Rejected background}
\label{sect:app_sac_bkg}

We define $B$ as the background valid event counts accumulated over the region of interest, quadrant or full detector, per unit frame, averaged over many frames. We shall assume the background to be distributed uniformly over the region of interest. We define $P_{rb}$ as the probability that, in a given frame, a valid event produced by a cosmic ray is rejected by SAC, i.e., $P_{rb} \equiv r_b/t_b$, where $r_b$ is the number of rejected background events and $t_b$ the total number of background events in the frame. Two distinct terms contribute to $P_{rb}$:
\begin{enumerate}
\item $P_{ran}$, the probability that the valid background event fell within the exclusion circle(s) of cosmic ray tracks from one or more unrelated primaries; and
\item $P_{cor}$, the probability that the valid background event fell within the exclusion circle of a particle track associated with the primary that generated it.
\end{enumerate}
The first term has already been introduced when discussing signal loss (see Eq.\ref{eq:p_sl}, $P_{ran} = P_{sl} = A_R /A_T$), and it depends both on the exclusion circle and the frame time. The second term depends on the exclusion circle, but does not depend on the frame time; it may be thought of in a simple way as a sort of cumulative ``secondary spread function'', $P_{cor} \equiv P_{cor}(<r_e)$, encapsulating the spatial spread of secondary particles. Like the fractional signal loss, $P_{cor}$ depends on the specific region that is being considered, and it will differ when considering a single quadrant or the full detector. A derivation of $ P_{cor}$ for these two cases and for the limiting case of an infinite plane is provided in Appendix \ref{sect:app_pcor}.

Note that $P_{rb}$ cannot be simply written down as the sum of $P_{cor}$ and $P_{sl}$. Indeed, for long frame times, a secondary event may end up falling simultaneously within the exclusion circle of the primary that generated it and in that of one or more unrelated primaries. This can be accounted for by including in the sum a correction term that accounts for the double counting of events that belong to both categories, i.e.:
\begin{equation}
P_{rb} = P_{cor} + P_{sl} - P_{cor} \cdot P_{sl} \, ,
\end{equation}
where $P_{cor} \cdot P_{sl} $ is the probability that an event ends up falling simultaneously within the exclusion circle of the primary that generated it and that of one or more unrelated primaries. By rearranging some of the terms and using Eq.\ref{eq:p_sl}  we can rewrite the above equation in the form:
\begin{equation}
\label{eq:p_rb}
P_{rb} = ( 1 - P_{cor}) \cdot A_R /A_T + P_{cor} \, .
\end{equation}
Once $P_{rb} $ is known, the background  can be computed from the equation:
\begin{equation}
\label{eq:b}
B = (1 - P_{rb}) \cdot B_o \, ,
\end{equation}
where $B_o$ is the background when no SAC is applied. Substituting Eq.\ref{eq:p_rb} into Eq.\ref{eq:b} we find
\begin{equation}
\label{eq:b2}
B = \{1 - [( 1 - P_{cor}) \cdot A_R /A_T + P_{cor}]\} \cdot B_o \, .
\end{equation}

\begin{figure}[t]
\begin{center}
\includegraphics[width=4in]{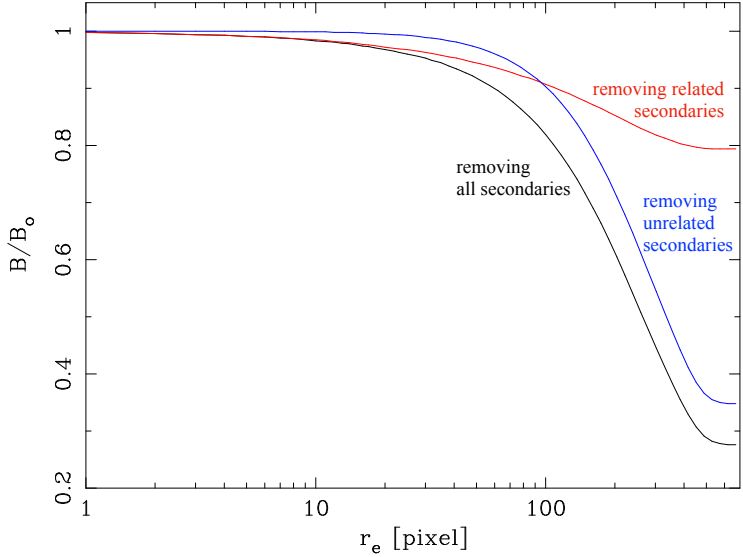}
\caption{Fractional background rejection as a function of exclusion radius for a 5 ms frame time and a 510$\times$510 grid, representing a single quadrant. We show in black the total fractional background; in red the fractional background if only secondaries that fall in the exclusion circle of generating primaries are removed; in blue the fractional background if only secondaries that fall in the exclusion circle of unrelated primaries are removed.}
\label{fig:bkg}
\end{center}
\end{figure}

In Fig.\ref{fig:bkg} we use Eq.\ref{eq:b2} to plot the fractional background, $B/B_o$, as a function of the exclusion radius, for a frame time of 5 ms. $P_{cor}$ has been derived from WFI simulations \cite{vonKienlin2018} as described in detail in Appendix \ref{sect:app_pcor}, and $A_R /A_T$ has been derived from Monte Carlo simulations as described in Section \ref{sect:app_sac_signal}.

Fig.\ref{fig:bkg} illustrates why $B$ is not a good SAC estimator.  Application of self anti-coincidence results in two different kinds of background reduction: a favorable one, associated with the removal of secondaries that fall in the exclusion circle of the primaries that generated them (Fig.\ref{fig:bkg} red curve); and an unfavorable one, associated with the removal of unrelated secondaries (Fig.\ref{fig:bkg} blue curve). Using only estimator $B$, we do not have a way of discriminating between the two.
%A better approach is to use the signal which tracks the unfavourable change and the SNR and S/N %which follow the favourable one.

\subsection{Signal-to-background ratio}
\label{sect:app_sac_SB}

The signal-to-background ratio, $S/B$, can be easily worked out from the equations for the signal and background, Eqs.\ref{eq:s} and \ref{eq:b2}. With a little algebra we find:
\begin{equation}
\label{eq:sb}
\frac{S}{B}  = \frac{S_o}{B_o} \cdot  \frac{1}{(1 - P_{cor})} \, .
\end{equation}
Interestingly, unlike $S$ and $B$, $S/B$ does not depend upon frame time but only on the exclusion radius through $P_{cor}$. Moreover, if we divide both sides of the equation by $S_o/B_o$, and define a re-normalized signal to background ratio
\begin{equation}
\frac{s}{b} \equiv \frac{S/S_o}{B/B_o} \, ,
\end{equation}
we derive a very general formula,
\begin{equation}
\label{eq:sbn}
\frac{s}{b}  =  \frac{1}{(1 - P_{cor})} \, ,
\end{equation}
which does not depend on the specific values of $S_o$ or $B_o$.

\begin{figure}[t]
\begin{center}
\includegraphics[width=4in]{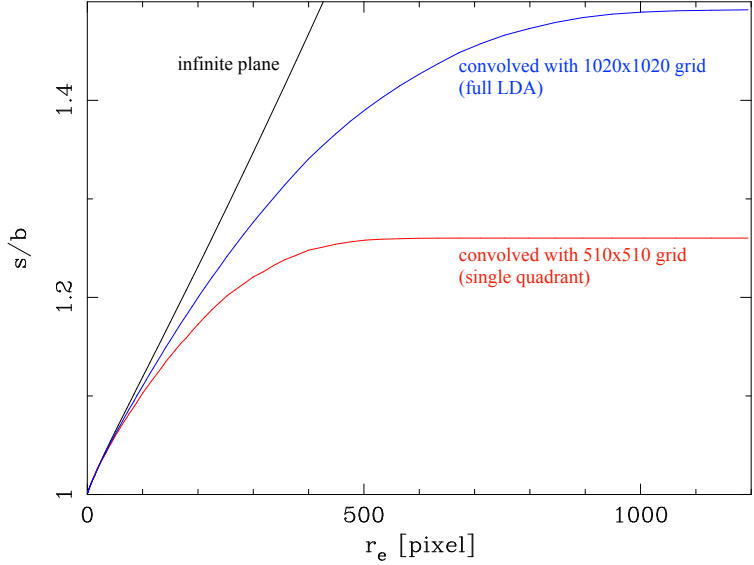}
\caption{Renormalized signal to background ratio. Different colors refer to different regions: black refers to an infinite plane, blue to a 1020$\times$1020 grid representing the full detector and red to a 510$\times$510 grid representing a single quadrant.}
\label{fig:sb}
\end{center}
\end{figure}

In Fig.\ref{fig:sb} we plot the renormalized signal to background ratio, where $P_{cor}$ has been derived from WFI simulations \cite{vonKienlin2018} as described in detail in Appendix \ref{sect:app_pcor}. As already pointed out, $S/B$ does not depend on area loss, and therefore on frame time; it depends only on  $P_{cor}(<r_e)$, i.e., the cumulative probability distribution that a secondary fall within a radius $r_e$ of its primary. $S/B$ improves steadily, reaching a maximum values of $\sim 20\%$ for rejection of a entire quadrant and $\sim 50\%$ for rejection of the full detector area. $S/B$ is a good indicator of the improvement afforded by SAC, however it does not provide a full description of its effects because it contains no information of the area loss.

\subsection{Signal-to-noise ratio}
\label{sect:app_sac_SNR}

The signal-to-noise ratio 
\begin{equation}
SNR \equiv \frac{S}{{(S+B)}^{1/2}} \cdot t^{1/2}\, ,
\end{equation}
where $t$ is the exposure time, can also be easily worked out from the equations for the signal and background, Eqs.\ref{eq:s} and \ref{eq:b2}. With a little algebra we find:
\begin{equation}
\label{eq:snr}
{SNR}  = \frac{(1 - A_R /A_T)^{1/2} \cdot S_o}{[S_o +  (1 - P_{cor}) \cdot B_o]^{1/2}}\cdot t^{1/2} \, .
\end{equation}
Interestingly, like $S$ and $B$, and unlike $S/B$, $SNR$ does depend upon frame time through $A_R$ as well as on the exclusion radius through $P_{cor}$.  In the background dominated regime, $B \gg S$, where
\begin{equation}
{SNR}  = \frac{S}{B^{1/2}} \cdot t^{1/2} \, ,
\end{equation} 
Eq.\ref{eq:snr} reduces to:
\begin{equation}
{SNR}  = \frac{S_o}{B_o^{1/2}} \cdot \frac{ (1 - A_R /A_T)^{1/2}}{(1 - P_{cor})^{1/2}} \cdot t^{1/2} \, .
\end{equation}
If we divide both sides of this equation by $S_o/\sqrt{B_o}\cdot t^{1/2}$, and define a re-normalized signal to noise ratio:
\begin{equation}
\label{eq:snr_sbsqrt}
{snr} \equiv \frac{S/S_o}{(B/B_o)^{1/2}} \, ,
\end{equation}
we derive a very general formula,
\begin{equation}
\label{eq:snrn}
{snr}  =  \frac{(1 - A_R /A_T)^{1/2}}{(1 - P_{cor})^{1/2}} \, ,
\end{equation}
which does not depend on the specific values of $S_o$ or $B_o$.

\begin{figure}[t]
\begin{center}
\includegraphics[width=\linewidth]{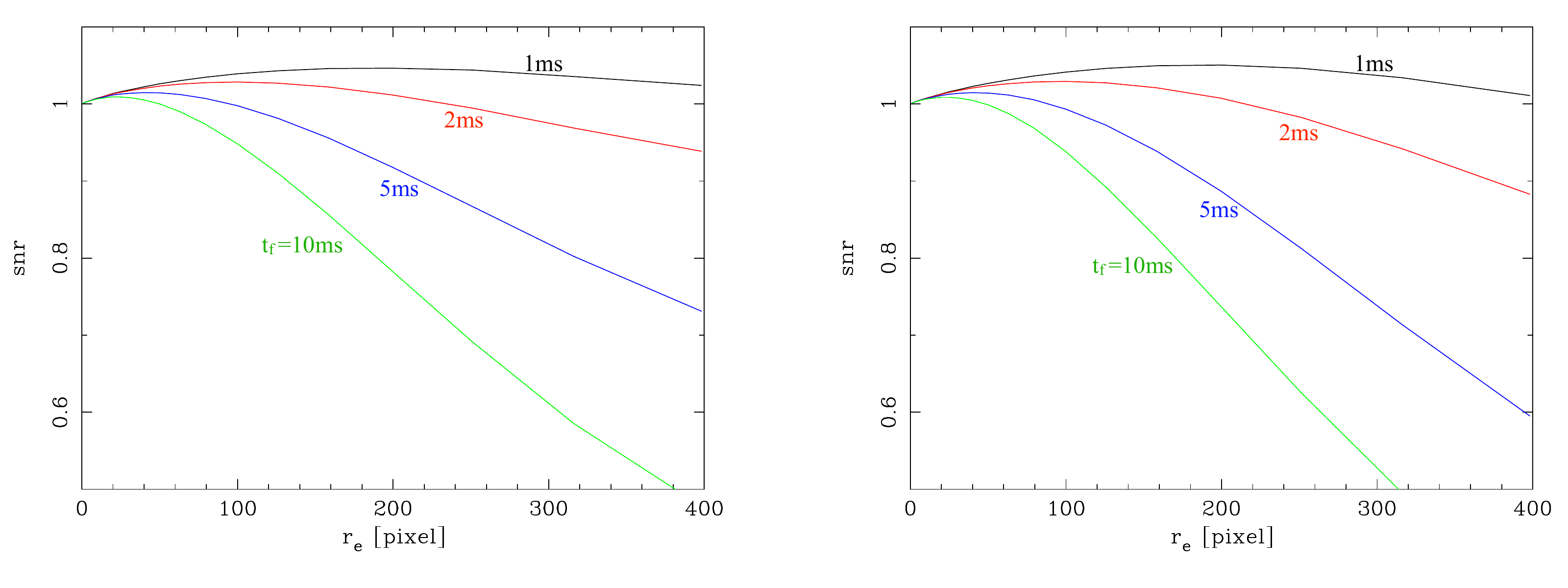}
\caption{Renormalized signal-to-noise ratio in the background-dominated regime as a function of exclusion radius for four different values of the frame time, as indicated by different colors. The left panel, for a 510$\times$510 grid, represents a single WFI LDA quadrant; the right panel, for a 1020$\times$1020 grid, represents the full detector.}
\label{fig:app_snr}
\end{center}
\end{figure}

As already pointed out, like $B$, $snr$ depends on area loss and on $P_{cor}$, and so it is a mixed estimator. However, unlike $B$, it can be of use by informing us about the exclusion radius that maximizes the signal-to-noise ratio. By looking at Fig.\ref{fig:app_snr}, left or right panel, we see that for a given choice of frame time, $snr$ peaks at specific values of the exclusion radius. For $t_f=10$ ms, maximum $snr$ is reached at $r_e = 30$ pixels; as we reduce the frame time, the peak moves to larger exclusion radius. For $t_f=1$ ms the $snr$ attains its peak value at $r_e \sim 200$ pixel. However, in all instances the maximum improvement on the $snr$ is less than 5\% with respect to the no-SAC case, so this is not a particularly significant improvement. In simpler words, application of Self Anti-Coincidence, be it to a single quadrant (Fig.\ref{fig:app_snr} left panel) or the full detector (Fig.\ref{fig:app_snr} right panel), does not improve the \textit{statistical} quality of our data in a significant way.

It is worth pointing out that, although the present result has been derived in the background dominated regime, $B \gg S$, it applies to all regimes. Indeed, as we can see in Eq.\ref{eq:snr}, when the signal $S$ is larger or comparable to the background $B$, the noise term becomes less sensitive to the value of the background and signal-to-noise improvements associated with background reduction become even less significant than in the background dominated regime.

\subsection{Effects of a Rolling Shutter}
\label{sect:app_rolling_shutter}

\begin{figure}[p]
\begin{center}
\includegraphics[width=5.0in]{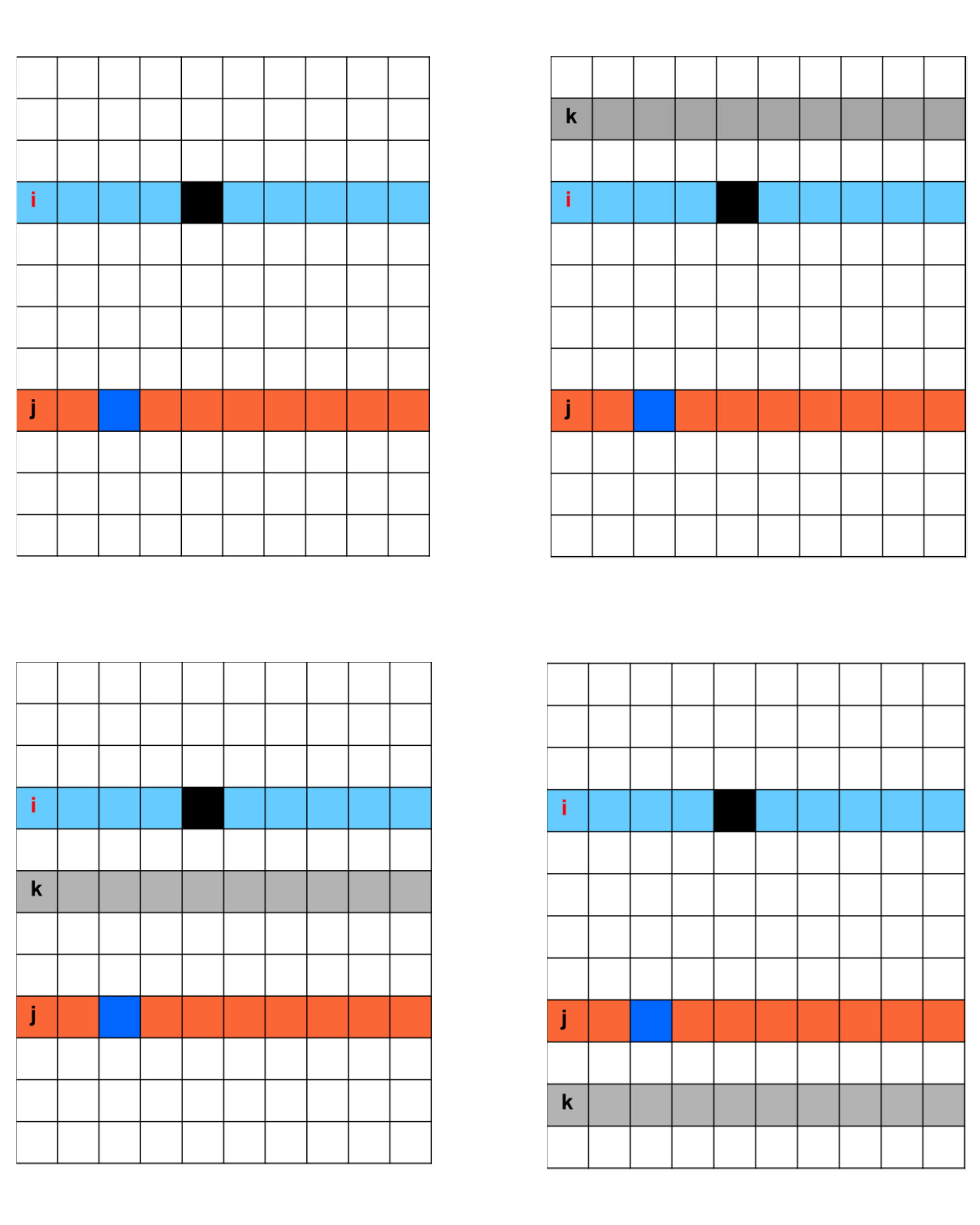}
\caption{Schematic representation of how primary and secondary events are read out, note that the rolling shutter moves from top to bottom. Top left panel: the primary (black square) hits on row $i$, indicated in light blue, and the secondary (blue square) impacts on row $j$ highlighted in orange. Top right panel: primary and secondary are placed as in the top left panel, also shown in gray is row $k$, which is being read out when the events hit the sensor. In this case, row $k$ is read out before rows $i$ and $j$ and both events are read out in the same frame. Bottom left panel: same as top right panel, however in this case row $k$ is placed between rows $i$ and $j$, this leads to the primary being read out one frame after the secondary. Bottom right panel: same as top right however, in this case, the rolling shutter goes through row $k$ after rows $i$ and $j$, this leads to both events being read out in the frame after the one depicted here.}
\label{fig:app_rolling_shutter_schematic}
\end{center}
\end{figure}

The above calculations have been performed tacitly assuming that all rows in a frame are read out simultaneously; this is true for many detectors, but not for the WFI LDA, which is operated in a `rolling shutter' mode. This is a read out mode where rows are continuously read out one after the other, frame after frame, and it is driven by technical limitations of the spacecraft power supply and thermal control\cite{Meidinger2017}. It also means that different rows are read out at different times and that, in some instances, a primary event and its secondary may end up recorded in different frames. In Figure \ref{fig:app_rolling_shutter_schematic}, we provide a schematic representation of how primary and secondary events are read out, with the rolling shutter moving from top to bottom. In the top left panel we show the primary (black square) on row $i$, indicated in light blue, and the secondary (blue square) on row $j$, highlighted in orange. Rows are numbered from top to bottom in accordance with the way the rolling shutter is operated, thus in the current example $i<j$. In the other three panels we also show row $k$ which is being  read out when primary and secondary hit the sensor. In the top right panel, $k<i<j$, and thus row $k$ is read out before rows $i$ and $j$ and both events end up in the same frame. In the bottom left panel, $i < k < j$, and the primary is read out one frame after the secondary. In the bottom right panel, $i < j < k$, and both events are read out in the next frame. Note that the velocity at which particles propagate in and around the detector is much larger than that at which the shutter is operated. Thus, within the current assessment, we can safely assume that primary and secondary impact the detector at the same time. 

Through a representation similar to the one presented in Figure \ref{fig:app_rolling_shutter_schematic}, it is easy to show that in the case where the primary lands on a higher row than the secondary, $j < i$, we can distinguish between three possible cases: 1) $k < j < i$, both events are read out in the same frame; 2) $j < k < i$, the secondary is read out one frame after the primary; and 3) $j < i < k$, both events are read out in the same frame. Finally, if primary and secondary hit on the same row, $i = j$, the two will be read out in the same frame.

In summary, for any given value of $i$ the secondary is read out during one of two frames; which of the two depends on the row $k$ that is being read out when primary and secondary impinge on the detector. (There is one minor exception to this rule: when $i = j$, the two events are read out in the same frame for any value of $k$.) If ($k<i$ and $k<j$) or ($k>i$ and $k>j$) the secondary will be read out in the same frame as the primary; if ($i<k<j$) or ($j<k<i$) the secondary and primary will be read out in different frames. Probabilities for the above cases can be easily computed. We make use of the following definitions: $P_{=}$ is the probability of primary and secondary being read out in same frame; $P_{\neq}$ is the probability of primary and secondary being read out in different frames; and $n_r$ is the number of rows. By requiring that the sum of all probabilities be unity we derive
\begin{equation}
\label{eq:roll1}
P_{=} + P_{\neq} = 1 \, ,
\end{equation}
and by noting that the probability of secondary and primary to be read out in different frames must be proportional to the number of rows between $i$ and $j$, 
\begin{equation}
\label{eq:roll2}
P_{\neq} = \frac{|i-j|}{n_r} \, .
\end{equation}
Finally, by combining Eq.\ref{eq:roll1} and \ref{eq:roll2} we derive:
\begin{equation}
\label{eq:roll3}
P_{=} = \frac{n_r - |i-j|}{n_r} \, .
\end{equation}

\begin{figure}[p]
\begin{center}
\includegraphics[width=\linewidth]{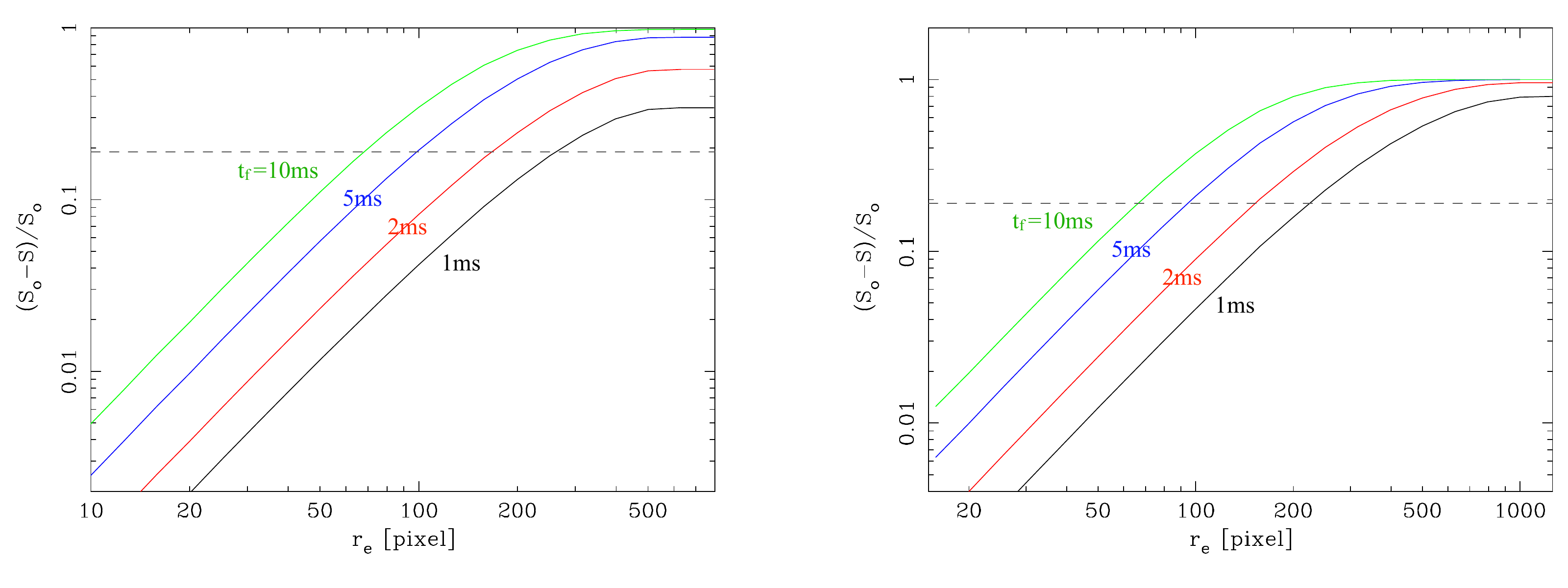}
\caption{Fractional signal loss as a function of exclusion radius for four different values of the frame time, as indicated by different colors. The left panel, for a 510$\times$510 grid, represents a single WFI LDA quadrant; the right panel, for a 1020$\times$1020 grid, represents the full LDA FOV. The effect of the rolling shutter has been included following the ``maximal exclusion'' option.}
\label{fig:app_rolling_sig_loss}
\end{center}
\end{figure}

From our analysis, we have determined that, except for the case where the secondary falls in the same row as the primary, the secondary can always be found in one of two frames. The question then is how to incorporate this information into our SAC calculations. We can consider two limiting approaches: 1) ``minimal exclusion'', throwing away only rows from the frame where the primary is detected; in this case the area loss term $A_R/A_T$ is unchanged, but $P_{cor}$, the probability that the secondary falls within the exclusion circle of the primary that generated it, will be significantly diminished, by up to a factor of two; and 2) ``maximal exclusion'', removing rows from both frames; in this case it is $P_{cor}$ that remains unchanged while the $A_R/A_T$ increases, again by a factor of about two. A wide range of intermediate solutions could be also considered. One could exclude rows from one frame only for rows that are close to the row in which the primary is located (rows with high $P_=$), and exclude rows from both frames for other rows (with low $P_=$). In the current work we shall use ``maximal exclusion'' as the most conservative background reduction case. In Figure \ref{fig:app_rolling_sig_loss}, we show the fractional signal loss as a function of exclusion radius, as shown in Figure \ref{fig:sig_loss} but with the inclusion of the rolling shutter effect.

\begin{figure}[p]
\begin{center}
\includegraphics[width=\linewidth]{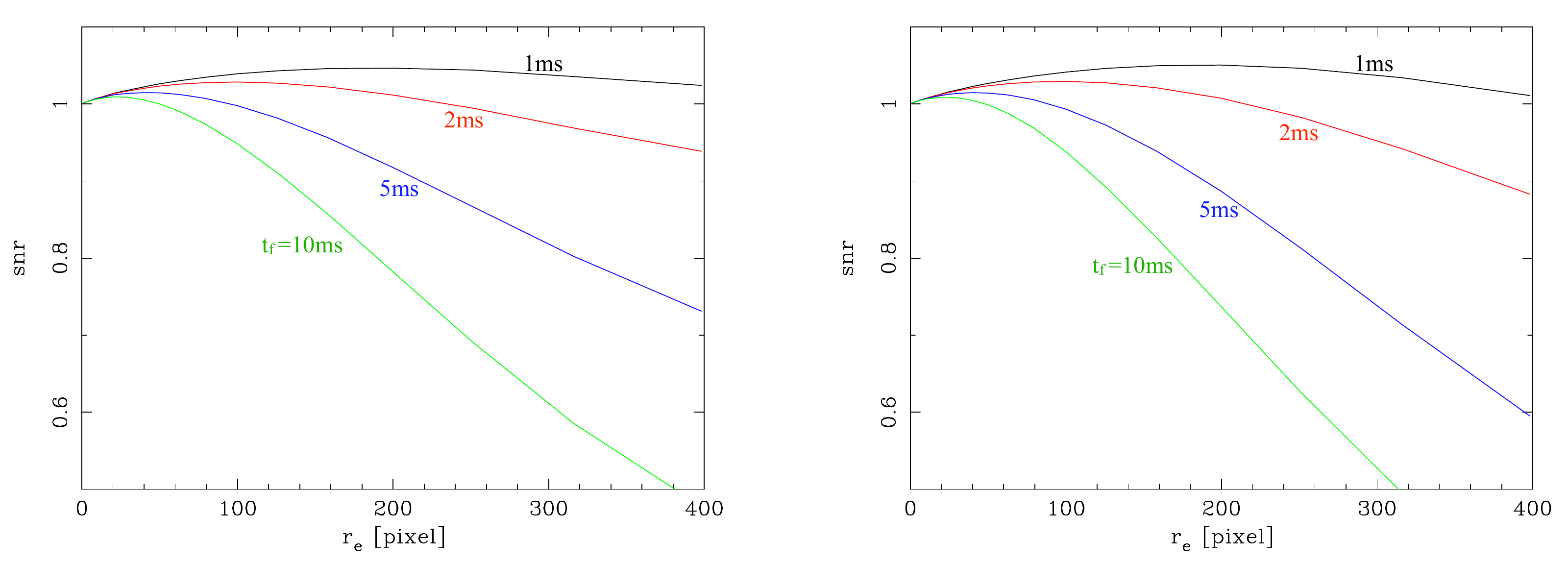}
\caption{Renormalized signal-to-noise ratio in the background-dominated regime as a function of exclusion radius for four different values of the frame time, as indicated by different colors. The left panel, for a 510$\times$510 grid, represents a single WFI LDA quadrant; the right panel, for a 1020$\times$1020 grid, represents the full detector. The effect of the rolling shutter has been included following the ``maximal exclusion'' option.}
\label{fig:app_rolling_snr}
\end{center}
\end{figure}

In the linear regime, i.e., $(S_0-S)/S_0 \ll 1$, the increase in signal loss is about a factor of two. However, as we move to larger exclusion radii, overlaps between excluded regions become more frequent and the increase in signal loss becomes smaller. By adopting the ``maximal exclusion'' option, the signal-to-background ratio $s/b$ remains unchanged because $P_{cor}$ remains unchanged. Conversely, since the signal-to-noise ratio depends on the area loss, it will be affected by the rolling shutter. This is shown in Figure \ref{fig:app_rolling_snr}, where we show the same plots as in Figure \ref{fig:app_snr} with the rolling shutter effect included.

%%%%%%%%%%%%%%%%%%%%%%%%%%%%%%%%%%%%%%%%%%%%%%%%%%%%%%%%%%%%%%%
% Appendix 2
%%%%%%%%%%%%%%%%%%%%%%%%%%%%%%%%%%%%%%%%%%%%%%%%%%%%%%%%%%%%%%%
\section{Computing the secondary distribution function for WFI}
\label{sect:app_pcor}

As pointed out in Section \ref{sect:app_sac_bkg}, the probability that a secondary is detected within a certain radius of the primary generating it, $P_{cor}(<r_e)$, depends on the specific region that is being considered, in other words it will differ when considering a single quadrant or the full detector. To derive $P_{cor}(<r_e)$ for the WFI we use the following procedure. We define the probability, $P^{\infty}_{cor}(<r_e)$, for the ideal case of an infinite plane as a parametric function of the form:
\begin{equation}
P^{\infty}_{cor}(<r_e) = 2/\pi \arctan[(r_e/r_\ast)^\alpha] \, ,
\end{equation}
where the free parameters are the scale radius $r_\ast$ and the slope of the power law $\alpha$. We insert trial values for $r_\ast$ and $\alpha$ and perform Monte Carlo simulations to compute from $P^{\infty}_{cor}(<r_e)$ the probabilities for a single quadrant, $P^{q}_{cor}(<r_e)$, and the full detector, $P^{d}_{cor}(<r_e)$. We then use Eq.\ref{eq:sbn} to compute the normalized signal-to-background ratio for a single quadrant, $(s/b)_q$,  and the full detector, $(s/b)_d$ and compare these with estimates based on detailed Geant4 simulation of the WFI. For this exercise we use previously published data from a different set of Geant4 simulations \cite{vonKienlin2018} for the single quadrant and a value provided to the Athena WFI Consortium (T.~Eraerds, private communication) for the full detector. We then iterate the procedure until $(s/b)_q$ and $(s/b)_d$ adequately reproduce the estimates based on Geant4 simulations.

\begin{figure}[t]
\begin{center}
\includegraphics[width=4in]{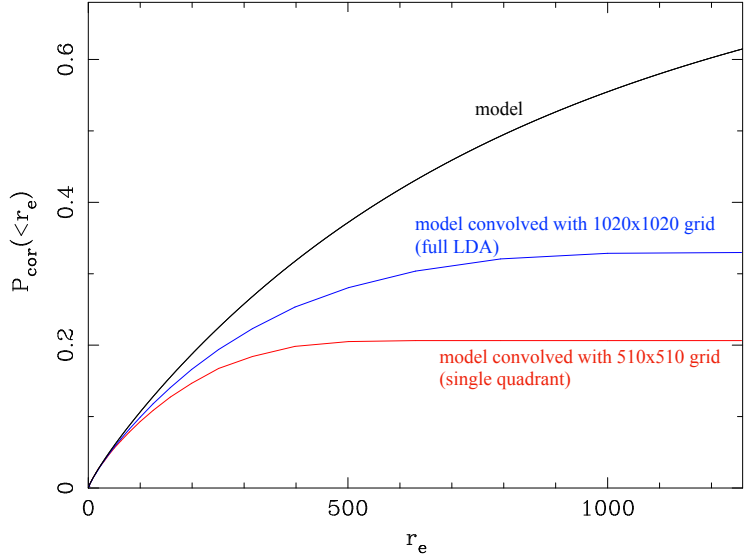}
\caption{Probability that a secondary fall within $r_e$ of its primary. The black, blue, and red curves refer respectively to: an infinite plane, a 1020$\times$1020 grid (detector), and a 510$\times$510 grid (quadrant).}
\label{fig:prob_cor}
\end{center}
\end{figure}

In Fig.\ref{fig:prob_cor} we show the probability distributions  $P^{\infty}_{cor}(<r_e)$, $P^{q}_{cor}(<r_e)$ and $P^{d}_{cor}(<r_e)$ that have resulted from the procedure we have just described. Note how, for large radii, the three curves converge to different values:  0.21 for $P^{q}_{cor}(<r_e)$, 0.32 for $P^{d}_{cor}(<r_e)$ and 1, by construction, for $P^{\infty}_{cor}(<r_e)$.

\begin{figure}[t]
\begin{center}
\includegraphics[width=\linewidth]{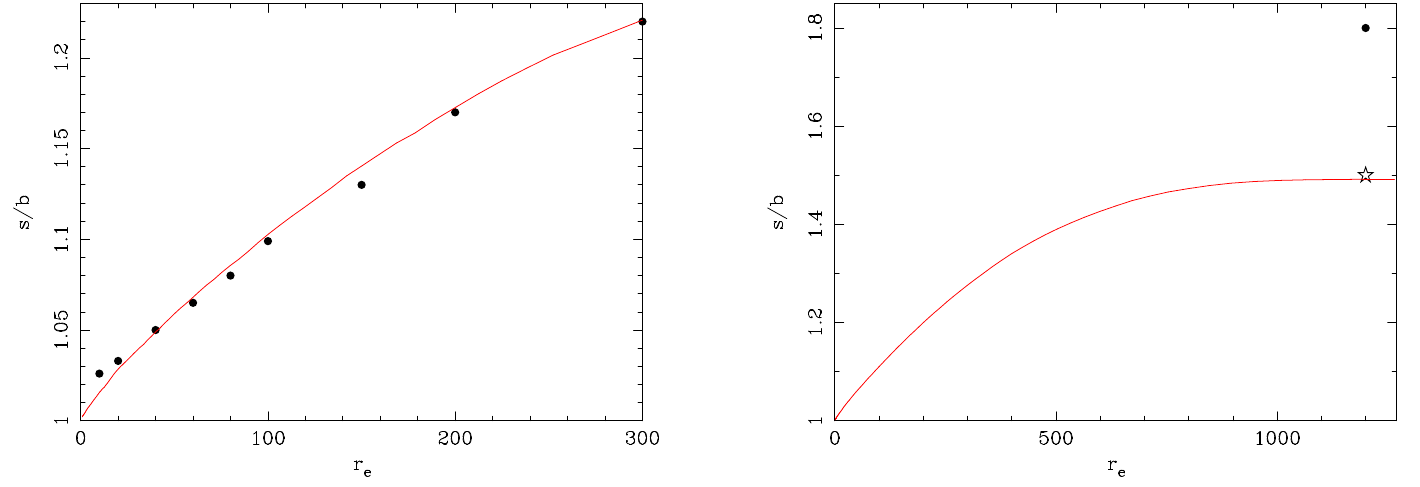}
\caption{Renormalized signal-to-background ratio for 510$\times$510 grid (quadrant, left panel) and 1020$\times$1020 grid (detector, right panel) vs. exclusion radius. Symbols come from Geant4 simulations, curves are computed through Eq.\ref{eq:sbn} from $P^{q}_{cor}(<r_e)$ for the left panel and $P^{d}_{cor}(<r_e)$ for the right panel.}
\label{fig:sb_fit}
\end{center}
\end{figure}

In Fig.\ref{fig:sb_fit} we show the normalized signal-to-background ratio for a single quadrant, $(s/b)_q$, and the full detector, $(s/b)_d$ and compare these with estimates based on Geant4 simulations of the WFI. The careful reader may note that, while in the case of $(s/b)_q$, left panel, the model fits the data points reasonably well, for $(s/b)_d$, right panel, we have a point and a star which are respectively well above and in agreement with the model. The point comes from a presentation at a WFI consortium meeting where results for both a single quadrant and the full detector were shown; the star comes from a rescaling of the point with the ratio of the result for the quadrant presented at the same meeting divided by the more recent estimate shown in the left panel of Fig.\ref{fig:sb_fit}.

\end{document}